\documentclass[12pt]{iopartamsmath}

\usepackage{graphicx}
\usepackage{amssymb}
\usepackage[utf8]{inputenc}
\usepackage{amsmath}
\usepackage{subcaption}
\usepackage{verbatim}
\usepackage{geometry}
\usepackage{longtable}
\usepackage[square, numbers]{natbib}
\usepackage{pdflscape}
\usepackage{textgreek}
\usepackage{ulem}

\usepackage{color}

\begin{document}

\title{Wavelength modulation laser spectroscopy of N$_2$O at 17~\textmu m}
\author{Y. Wang$^1$, J. Rodewald$^1$, O. Lopez$^2$, M. Manceau$^2$, B. Darqui\'e$^2$, B. E. Sauer$^1$, M. R. Tarbutt$^1$\footnote{E-mail: m.tarbutt@imperial.ac.uk}}

\address{$^1$Center for Cold Matter, Imperial College London, London, SW7 2AZ, United Kingdom}
\address{$^2$Laboratoire de Physique des Lasers, CNRS, Universit\'e Sorbonne Paris Nord, 93430 Villetaneuse, France}

\vspace{10pt}

\begin{abstract}
Using a mid-infrared quantum cascade laser and wavelength modulation absorption spectroscopy, we measure the frequencies of ro-vibrational transitions of N$_2$O in the 17 $\mu$m region with uncertainties below 5 MHz. These lines, corresponding to the bending mode of the molecule, can be used for calibration of spectrometers in this spectral region. We present a model for the lineshapes of absorption features in wavelength modulation spectroscopy that takes into account Doppler broadening, collisional broadening, saturation of the absorption, and lineshape distortion due to frequency and intensity modulation. Combining our data with previous measurements, we provide a set of spectroscopic parameters for several vibrational states of N$_2$O. The lines measured here fall in the same spectral region as a mid-infrared frequency reference that we are currently developing using trapped, ultracold molecules. With such a frequency reference, the spectroscopic methods demonstrated here have the potential to improve frequency calibration in this part of the spectrum.
\end{abstract}

\section{Introduction}
\label{introduction}

The mid-infrared is an important spectral region with applications in trace gas detection~\cite{Kosterev2008, Galli2016, MartinMateos2017}, atmospheric monitoring~\cite{Daghestani2014,Robinson2021}, high resolution spectroscopy~\cite{Bielsa2008, Borri2019, D'Ambrosio2019, Asselin2017}, frequency metrology~\cite{Sow2014,Hansen2015, Argence2015, Santagata2019, Tran2024}, measurements of fundamental constants~\cite{Mejri2015} and tests of fundamental physics~\cite{Cournol2019, Mudiayi2021, Barontini2022}. The wavelength region between 7 and 20~$\mu$m is known as the molecular fingerprint region, because every molecule has a unique spectrum in this region. However, the spectroscopic potential of this region has been difficult to exploit because (i) spectra are often crowded, requiring high resolution, (ii) laser technology in many parts of this region is often poorly developed or non-existent, (iii) optical elements often absorb strongly, especially beyond 15~$\mu$m,  (iv) there are not many absolute frequency standards in this spectral region, and (v) calibration of the frequency axis is difficult. A new generation of quantum cascade lasers at wavelengths beyond 10~$\mu$m is helping to solve some of these problems, see e.g. \cite{Van2019}. Here, we use a quantum cascade laser to demonstrate laser spectroscopy at 17~$\mu$m with an accuracy of a few MHz.

Nitrous oxide, N$_2$O, is a common gas that is often used for frequency calibration in the mid-infrared~\cite{Maki1992}. Environmental monitoring of this gas is important since it is the dominant anthropogenic cause of ozone depletion and is a potent greenhouse gas whose atmospheric concentration has been rising for decades~\citep{Ravishankara2009}. The fundamental vibrational frequency of the bending ($\nu_2$) mode of N$_2$O lies near 17~$\mu$m, and the gas has a forest of strong absorption lines in this spectral region. 

There exist two large spectroscopic databases providing comprehensive line lists for N$_2$O throughout the mid-infrared. The first is the HITRAN database~\cite{Gordon2022} which gives a line list for N$_2$O based in part on the data provided by Toth~\cite{Toth2004}, which is calculated from spectroscopic constants derived from measurements by the same author. The second, which we call the NIST database, is described in Ref.~\cite{Maki1992} and supported by line lists provided in tabular form and online~\cite{NIST118}. The line list is built from spectroscopic constants derived from a wide range of data sources including microwave spectroscopy, Fourier transform spectroscopy and heterodyne frequency measurements. The heterodyne measurements are described in \cite{Whitford1975} and a series of papers from NIST that include Refs.~\cite{Pollock1984,Wells1985,Wells1985_a,Zink1987,Vanek1989}. They provide absolute frequencies for a selection of lines which are denoted as recommended calibration lines in the NIST database. In the region around 17~$\mu$m, the accuracy of these calibration lines reaches 3~MHz. We note that the frequencies of lines in the $\nu_2$ bands were not measured directly, but instead derived through a combination of frequency measurements of other bands at shorter wavelengths, e.g.~\cite{Vanek1989}.

There has been very little prior laser spectroscopy of the $\nu_2$ band of N$_2$O. In earlier work~\cite{Reisfeld1979, Baldacchini1992, Baldacchini1993} using a lead salt laser, a set of lines were measured with a relative precision of 30~MHz. Recently, a quantum cascade laser (QCL) operating at 17.2~$\mu$m was developed and its spectral properties were measured using absorption lines of N$_2$O~\cite{Manceau2023}. In this paper, we demonstrate wavelength modulation spectroscopy of N$_2$O using this QCL. We measure 165 lines in the small region between 580.27 and 582.48~cm$^{-1}$ and determine their frequencies. Because the method has high sensitivity and high dynamic range, we are able to observe lines from five different isotopologues and 12 different vibrational bands. 

At present, the Bureau International des Poids et Mesures recommends only two frequency standards in the mid-infrared, one at 3.39~$\mu$m based on CH$_4$ with a relative uncertainty of $3 \times 10^{-12}$, and the other at 10.3~$\mu$m based on OsO$_4$ with an uncertainty of $1.4 \times 10^{-13}$~\cite{Quinn2003}. Recent advances in cooling and controlling neutral molecules~\cite{Fitch2021b, Augenbraun2023} and molecular ions~\cite{Deiss2023} raise the prospect of a set of new frequency standards in the mid-infrared, based on the vibrational frequencies of trapped molecules, whose precision could match that available in the microwave and optical domains ($10^{-16}$ or better). These frequency standards could then be used to calibrate large sets of spectral lines in common gases such as N$_2$O. Suitable molecular clocks are currently being developed, motivated in part by their potential to search for varying fundamental constants~\cite{Leung2022,Barontini2022,Schiller2014,Chou2017,Sinhal2020}. The QCL used in the present work has been designed so that its frequency coincides with the fundamental vibrational frequency of CaF molecules, which can be cooled to microkelvin temperature~\cite{Caldwell2019} and can make an outstanding frequency standard~\cite{Barontini2022}. We consider the prospects of improving frequency calibration in this spectral region through the combination of an absolute frequency reference provided by CaF molecules, the set of strong and narrow absorption lines provided by N$_2$O and other gases, and the reliable source of laser light provided by QCL technology. 

\section{Methods}

\begin{figure*}[t]
    \centering
    \includegraphics[width=0.7\textwidth]{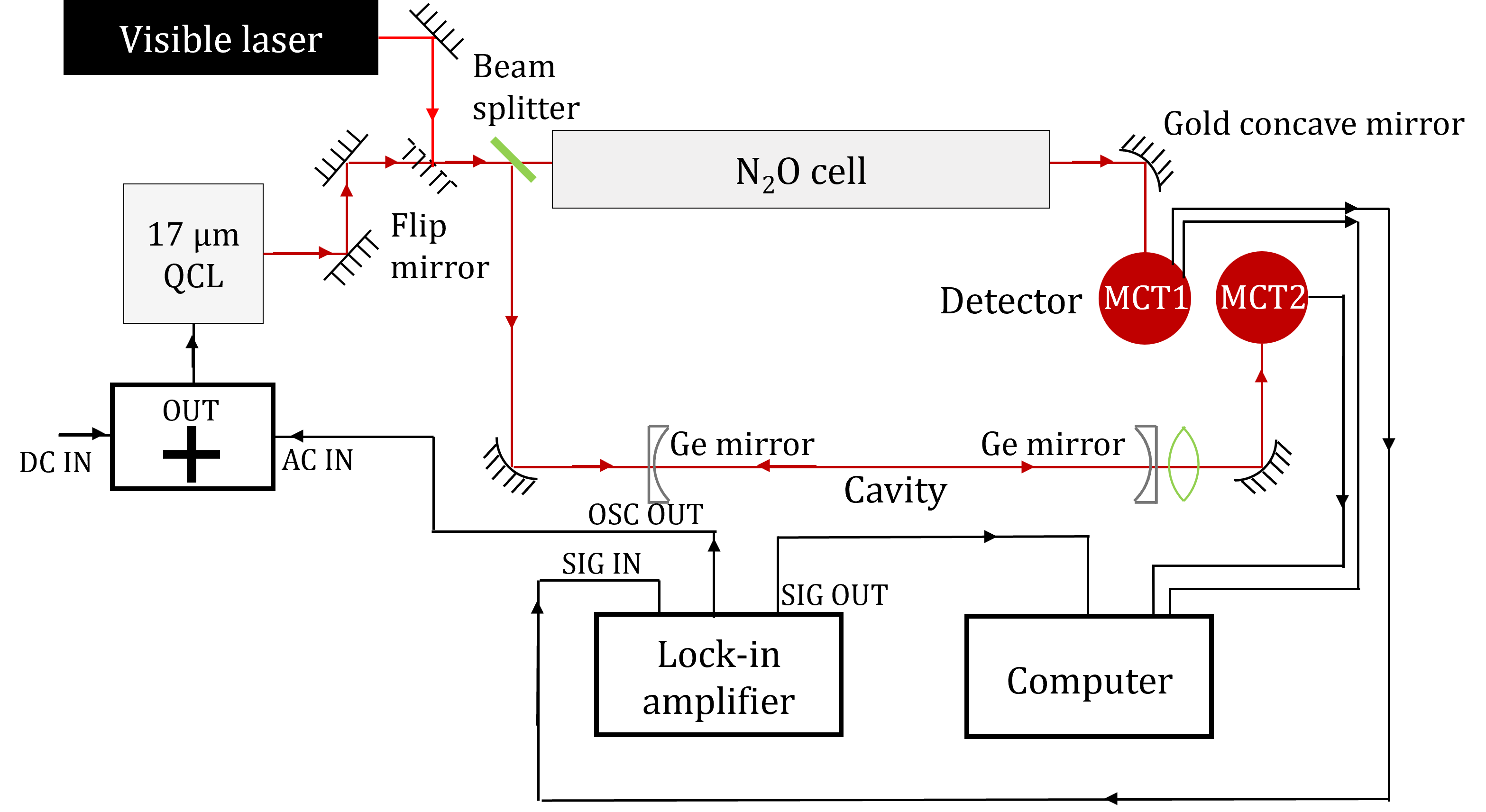}
    \caption{Schematic of the experimental setup. The 17 $\mu$m laser beam is divided into two parts, one passing through the N$_2$O cell for absorption spectroscopy, and the other through a cavity for frequency calibration. A visible laser is used as an alignment aid. The laser wavelength is modulated and the absorption signal at a harmonic of the modulation frequency is detected using a lock-in amplifier. OSC OUT is the modulation fed to the laser; SIG IN is the direct absortion signal; SIG OUT is the output of the lock-in amplifier.}
    \label{fig:17umbd}
\end{figure*}

Figure \ref{fig:17umbd} illustrates the experimental setup we use for wavelength modulation spectroscopy. The 17~\textmu m laser is a continuous wave distributed feedback QCL with an active region formed from InAs/AlSb~\cite{Manceau2023}. The frequency can be tuned by about 3~cm$^{-1}$ by changing the temperature and drive current. The temperature was set in the range 243 to 263~K and the current scanned between 500 and 570~mA. The light from the QCL is separated into two beams, each with a power of approximately 0.5~mW. 

The first beam, used for absorption spectroscopy, passes through a gas cell of length 20~cm containing about 350~Pa of N$_2$O, and then onto a liquid nitrogen cooled mercury cadmium telluride detector, MCT1, that measures its power. By applying an oscillating component to the current of the QCL, its frequency is modulated with $f_{\rm mod} =  2.22$~kHz and a modulation amplitude of roughly 20~MHz. This modulation amplitude was chosen to give the best signal-to-noise ratio without significantly broadening the lines. However, since the modulation amplitude is comparable to the typical linewidths of about 36~MHz, it is necessary to include the effects of the modulation in the lineshape model, as we discuss later and in \ref{appendix:WMS}. A lock-in amplifier detects the amplitude of the oscillating signal at MCT1 at either $f_{\rm mod}$ or $2f_{\rm mod}$. 

\begin{figure*}[t]
    \centering
    \includegraphics[width=0.8\textwidth]{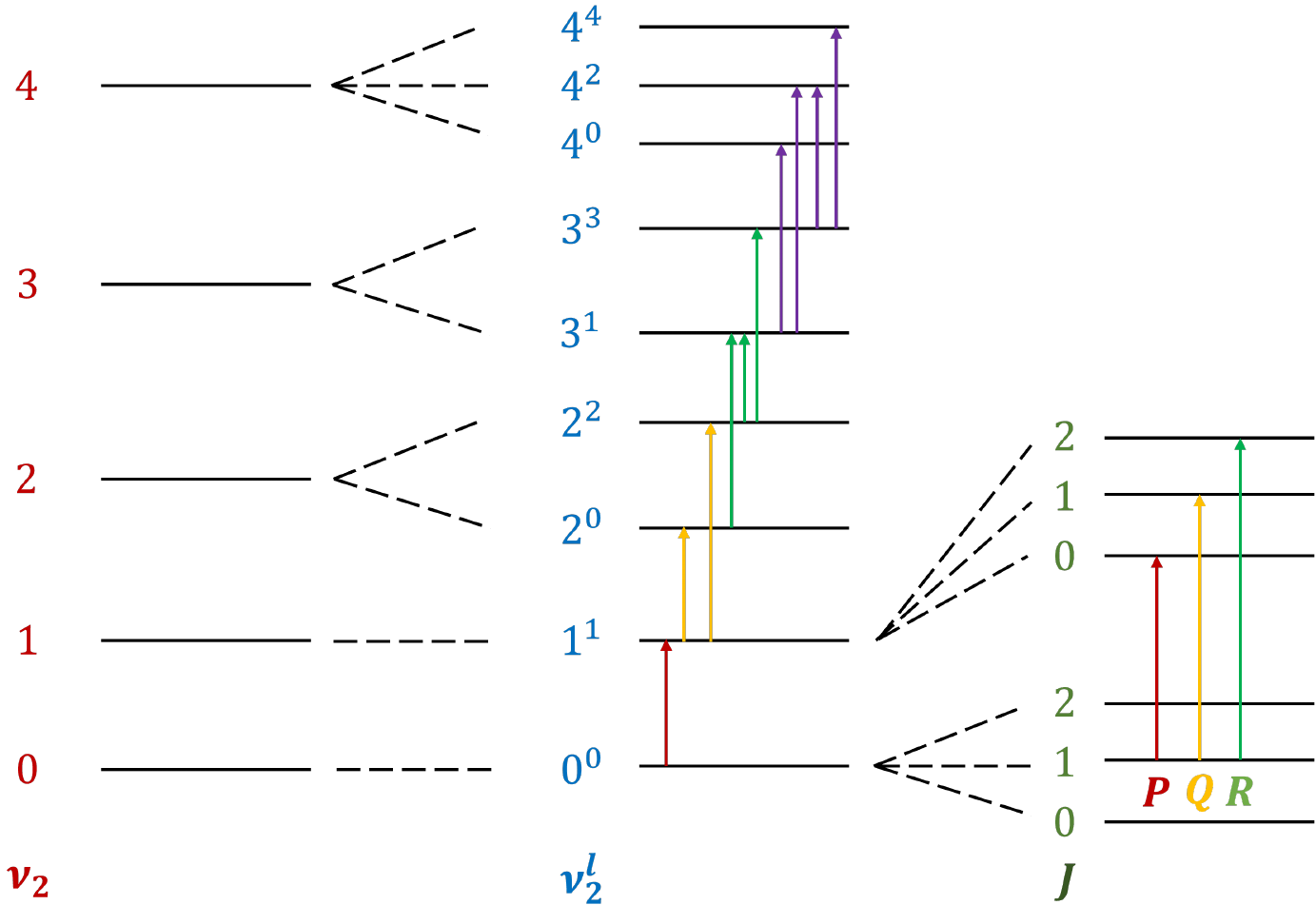}
    \caption{Relevant energy levels of N$_2$O and allowed transitions near 17~$\mu$m.  Vibrational states labelled by $\nu_2$ are split into levels according to the value of the bending mode angular momentum quantum number $l$, which is restricted to $0 \le l \le \nu_2$ and $l$ even (odd) when $\nu_2$ is even (odd). Each vibrational state has a ladder of rotational states labelled by the rotational angular momentum quantum number $J$ (we have only shown the first 3 rotational states for the lowest 2 values of $\nu_2$). All transitions we drive have $\Delta \nu_1 = \Delta \nu_3 = 0$, $\Delta \nu_2 = 1$, $\Delta l = \pm 1$. Rotational transitions are labelled as $P$, $Q$ and $R$ for $\Delta J = -1, 0 ,+1$ respectively. The quantum numbers $\nu_1, \nu_3$ are not shown since they do not change in the transitions.}
    \label{fig:levels}
\end{figure*}

The second beam, used to calibrate the frequency scale, passes through a confocal cavity and then onto a separate detector, MCT2. The cavity is formed from a pair of plano-concave lenses separated by 30~cm and made from germanium whose refractive index at 17.2~$\mu$m is $n=3.9997$, giving a reflectivity of 35.9\%. The cavity provides a set of frequency markers with a spacing of $\delta f = c/(4L) \approx 250$~MHz. The signals from MCT1, MCT2 and the lock-in amplifier are recorded simultaneously. The placements of all optical components are carefully adjusted to minimize etaloning effects from back reflections. 

\section{Results and analysis}

\begin{figure*}[p]
    \begin{subfigure}{\textwidth}
        \centering
        \includegraphics[width=\textwidth]{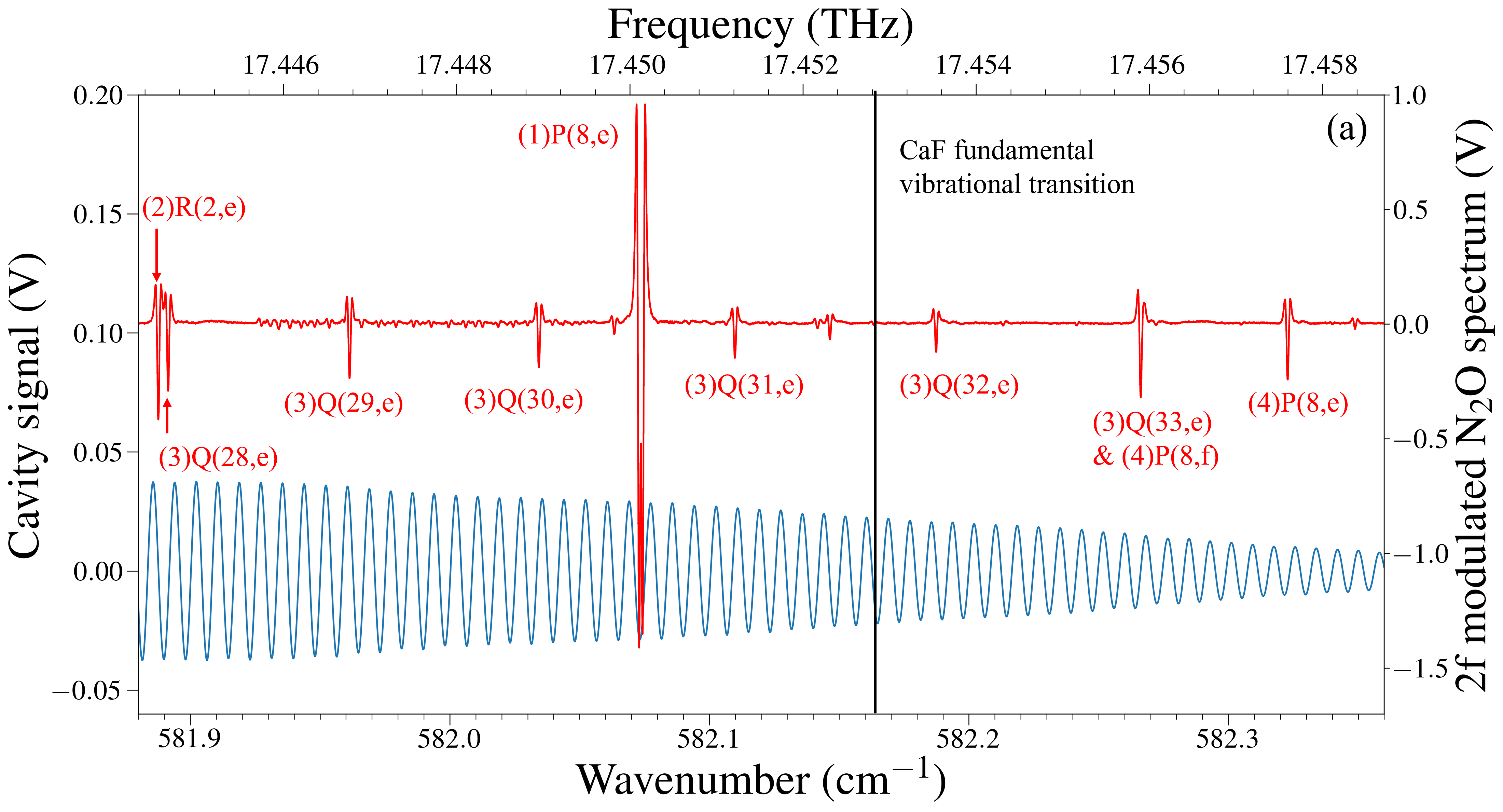}
    \end{subfigure}   
    \hfill
    \begin{subfigure}{\textwidth}
        \centering
        \includegraphics[width=\textwidth]{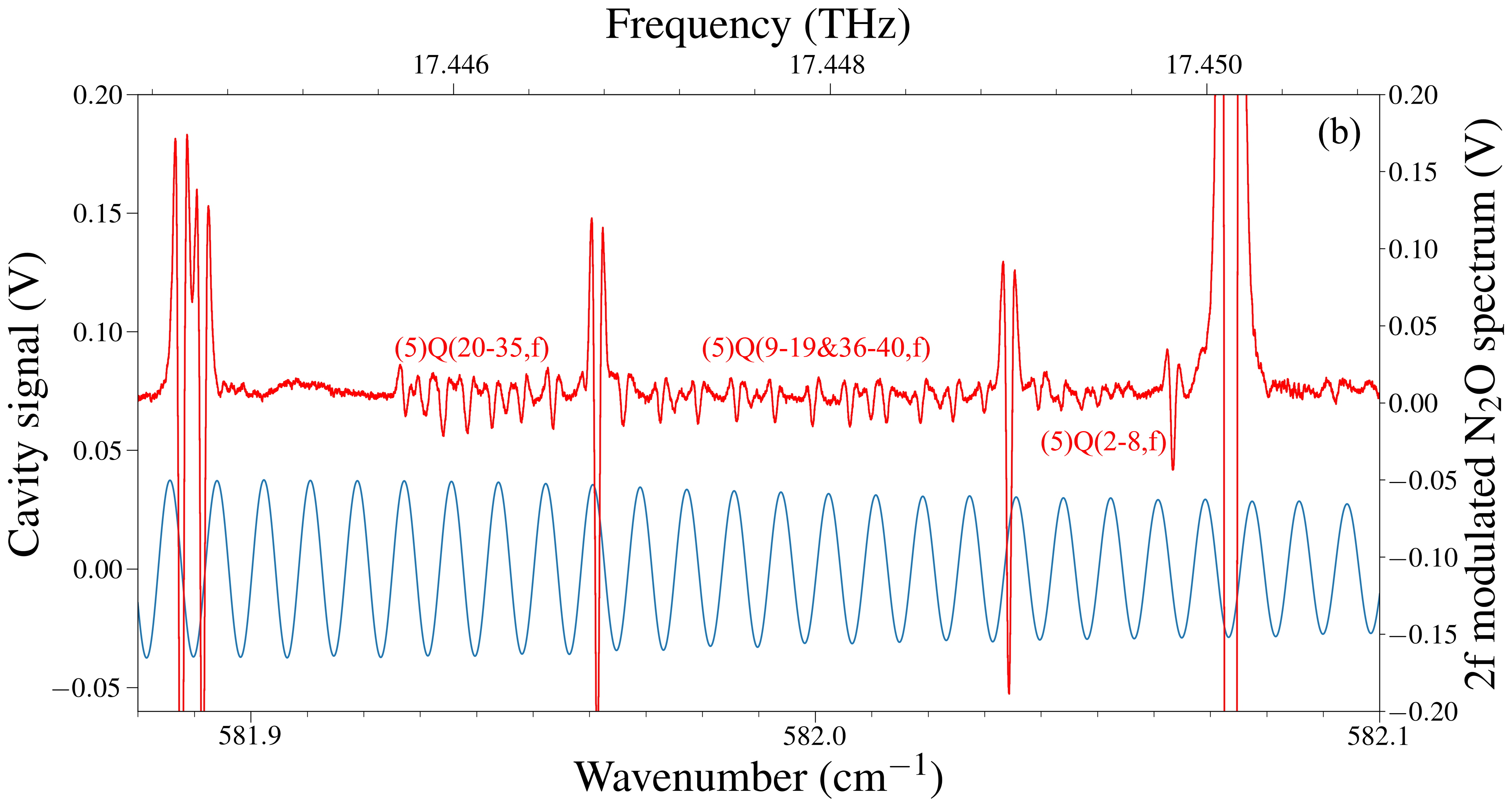}
    \end{subfigure}
       \caption{Examples of measured $^{14}$N$_2^{16}$O rovibrational transitions detected at $2f_{\rm mod}$ (red), together with cavity fringes (blue). (a) Part of spectrum spanning 0.48~cm$^{-1}$. Vertical black line indicates the frequency of the CaF $v'=1 \leftarrow v=0$ P(1) transition~\cite{Charron1995, Kaledin1999} (b) Expanded region spanning 0.22~cm$^{-1}$, showing many smaller lines. Transitions are labelled in the format `(band)P/Q/R(J,e/f)' with vibrational bands denoted: (1): $(01^10)-(000)$; (2): $(02^00)-(01^10)$; (3): $(03^10)-(02^00)$; (4): $(02^20)-(01^10)$ (5): $(04^20)-(03^10)$.}
    \label{fig:n2o_spectra_freq}
\end{figure*}

Figure \ref{fig:levels} illustrates the energy levels of N$_2$O relevant to this work. N$_2$O is a linear triatomic molecule with a $^1\Sigma$ ground state. Vibrational states are labelled as $(\nu_1\,\nu_2^l\,\nu_3)$ where $\nu_{1,2,3}$ are the quantum numbers of the symmetric stretch mode, the bending mode and the antisymmetric stretch mode respectively, and $l$ is the angular momentum associated with the bending mode. The labels $e$ and $f$ are used to denote the symmetry of the $l$-type doubling components following the standard convention\footnote{For integer $J$, levels with parity $(-1)^J$ are $e$ levels and levels with parity $-(-1)^J$ are $f$ levels. $Q$ transitions couple $e \leftrightarrow f$, whereas $P$ and $R$ transitions couple $e \leftrightarrow e$ or $f \leftrightarrow f$. Levels with $l=0$, where there is no $l$-doubling, are $e$ levels.}~\cite{Brown1975}. Rotational transitions that have $J'=J-1$, $J'=J$ and $J'=J+1$ are denoted as $P(J)$, $Q(J)$ and $R(J)$ respectively, where $J$ and $J'$ are the rotational angular momentum quantum numbers of the lower and upper levels. We focus on a region of the spectrum centred at 581.5~cm$^{-1}$ containing strong P lines of the $(01^10)-(00^00)$ band, weaker P lines of the $(02^20)-(01^10)$ band and Q lines of the $(03^10)-(02^00)$ band, and even weaker lines from several other bands. In total, we measure lines in the following 12 bands: $(01^10)-(00^00)$, $(02^00)-(01^10)$, $(02^20)-(01^10)$, $(03^10)-(02^00)$, $(03^10)-(02^20)$, $(03^30)-(02^20)$, $(04^00)-(03^10)$, $(04^20)-(03^10)$, $(04^20)-(03^30)$, $(11^10)-(10^00)$, $(12^00)-(11^10)$, $(12^20)-(11^10)$.

\subsection{Example data}

Figure \ref{fig:n2o_spectra_freq} shows examples of the data obtained. The red line is the N$_2$O absorption signal at $2f_{\rm mod}$ and the blue line is the cavity transmission. The laser frequency is scanned at a rate of approximately 180~MHz~s$^{-1}$ and the data presented are single scans. The frequency is controlled via the QCL current which also changes the power of the laser. That change is reflected in the amplitude of the cavity fringes as the laser is scanned, and its effect is included in our lineshape model (see \ref{appendix:WMS}). The frequency-dependence of the power is very close to linear so has little effect on the signal at  $2f_{\rm mod}$. 

Figure \ref{fig:n2o_spectra_freq}(a) shows a region of the spectrum spanning 0.48~cm$^{-1}$. The strongest line is the P(8) line of the $(01^10)-(00^00)$ band of $^{14}$N$_2^{16}$O. This is listed in the NIST database of recommended calibration lines and is one of the calibration lines used in this work. The other strong lines observed are mainly Q lines of the $(03^10)-(02^00)$ band in this isotopologue. Figure \ref{fig:n2o_spectra_freq}(b) is an expanded view of the small region between 581.88~cm$^{-1}$ and 582.09~cm$^{-1}$ where we see a dense spectrum of much weaker features which are Q lines in the $(04^20)-(03^10)$ band of $^{14}$N$_2^{16}$O. The signal-to-noise ratio for these weak features remains high. Indeed, we observe lines with strengths spanning 4 orders of magnitude (see below).

\subsection{Fitting to a model}

To fit to these data, we develop a model to describe the lineshapes. This model extends previous work~\citep{Arndt1965, Schilt2003} and accounts for Doppler broadening, collisional broadening, the effect of the modulation, saturation of the absorption at large optical depths, and effects due to the frequency dependence of the laser intensity. The model is important for determining accurate frequencies so we describe it in detail in \ref{appendix:WMS} and more briefly here. 

\begin{figure*}[t]
        \centering
        \includegraphics[width=\textwidth]{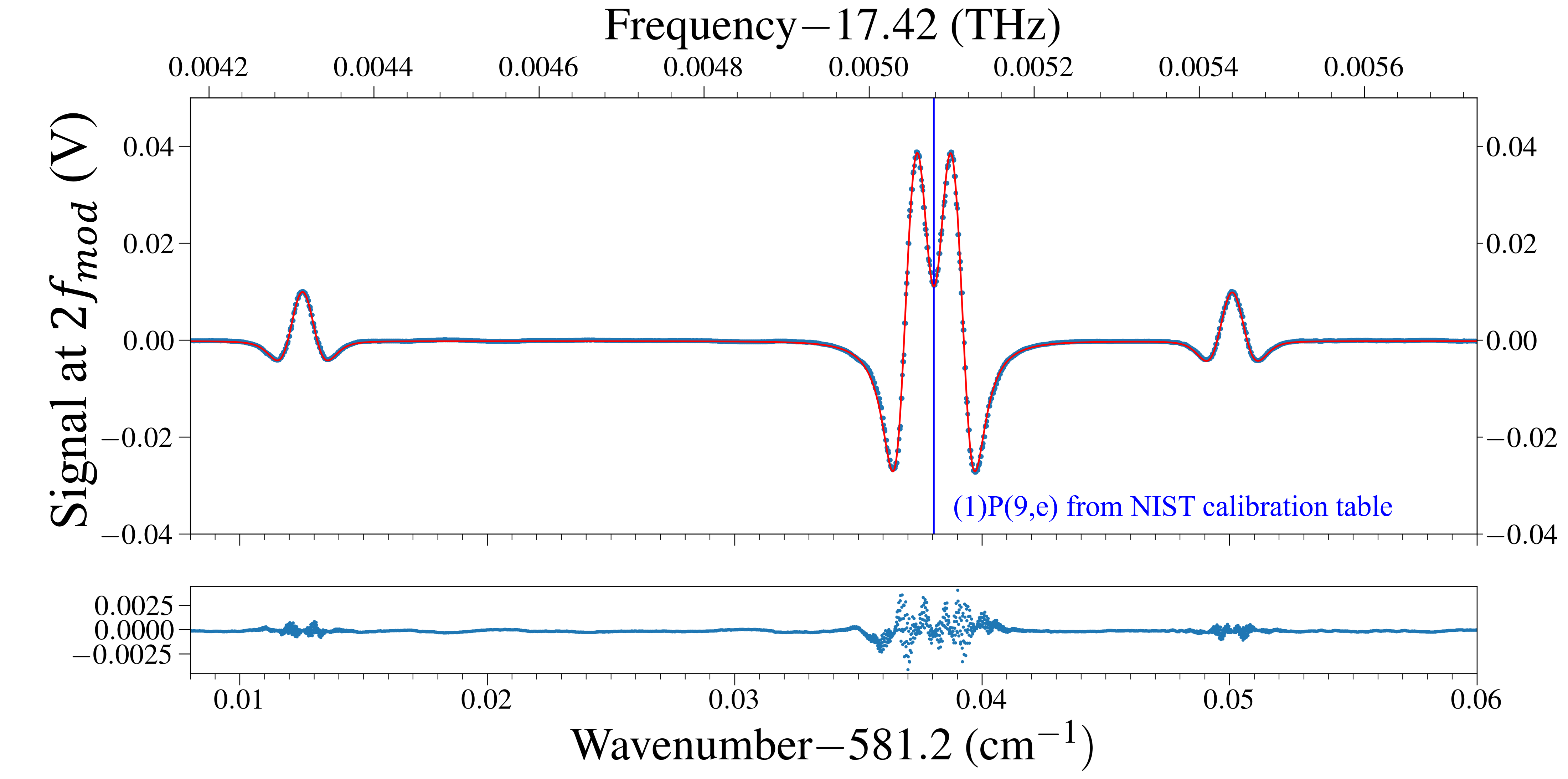}
    \caption{An example fit. Blue dots: absorption signal detected at $2f_{\rm mod}$; Red line: fit to model described by Eqs.~(\ref{eq:wms_lineshape}) and (\ref{eq:alpha}). Blue vertical line: (1)P(9,e) transition of $^{14}$N$_2^{16}$O used as a calibration line in this work. The bottom panel shows the fit residuals. Band labelling is the same as in figure \ref{fig:n2o_spectra_freq}.}
    \label{fig:n2o_calib_line}
\end{figure*}

The current of the QCL is modulated with angular frequency $\Omega = 2\pi f_{\rm mod}$, and this modulates both the wavelength and intensity of the light. The detuning of the light from the centre of an absorption feature, normalized to the linewidth of the absorption feature\footnote{This normalization is just for convenience; any measure of the linewidth can be used.}, is denoted by $\Delta$. It has two parts, $\Delta =  \delta - \delta_{\rm m}\cos(\Omega t)$, where $\delta$ is the detuning averaged over a modulation cycle and $\delta_{\rm m}$ is the modulation amplitude, both normalized to the linewidth. The laser intensity incident on the absorbing sample is $I_{\rm in}=I_0(1+p\Delta)$ where $I_0$ is the intensity at the centre of the absorption feature and we are assuming a linear dependence on $\Delta$ with proportionality constant $p$. We measure the intensity transmitted through the absorbing sample $I_{\rm out}$ and pick out the component of this signal oscillating at frequency $n\Omega$. As shown in \ref{appendix:WMS}, this component is
\begin{equation}
    \frac{I_{{\rm out},n}}{I_0} = 2\epsilon_n (-1)^n \int_{-\delta_{\rm m}}^{\delta_{\rm m}} \frac{\cos\left(n\cos^{-1}\left(\frac{\delta'}{\delta_{\rm m}}\right)\right)}{\sqrt{\delta_{\rm m}^2 - \delta'^2}}(1+p\delta-p\delta') e^{-\alpha(\delta-\delta')} d\delta',
    \label{eq:wms_lineshape}
\end{equation}
where $\epsilon_0 = 1$ and $\epsilon_n = 2$ for all $n>0$. The function $\alpha(\delta)$ is the optical depth and describes the frequency dependence of the absorption. In the present work, both Doppler broadening and collisional broadening are important, so we choose $\alpha(\delta)$ to be a Voigt profile,
\begin{equation}
    \alpha(\delta) = \int_{-\infty}^{\infty} G(\delta'', \sigma_1)L(\delta-\delta'',\sigma_2) d\delta''.
    \label{eq:alpha}
\end{equation}
Here $G(\delta,\sigma_1)$ is a Gaussian function with width parameter $\sigma_1$ and $L(\delta,\sigma_2)$ is a Lorentzian function with width parameter $\sigma_2$. To fit this model to our data, the integral in Eq.~(\ref{eq:wms_lineshape}) is calculated numerically with $\alpha$ determined at each value of $\delta'$ by numerical evaluation of the integral in Eq.~(\ref{eq:alpha}). This is used as the model in a standard non-linear fitting routine. In cases where there are overlapping lines, we still use Eq.~(\ref{eq:wms_lineshape}) but take $\alpha(\delta)$ to be a sum of Voigt profiles. An example of a fit to overlapping lines is given in \ref{appendix:WMS}.

Figure \ref{fig:n2o_calib_line} shows a small region of the spectrum around 581.2~cm$^{-1}$. The points are the data, and the line is a fit to the model described by Eqs.~(\ref{eq:wms_lineshape}) and (\ref{eq:alpha}).  The free parameters in the fit are the line centre, the peak optical depth and the Gaussian and Lorentzian widths. The value of $p$ is determined independently from the sloping background of the raw absorption data and is fixed in the fit. The parameter $\delta_{\rm m}$ is found for each segment of data from a fit to one of the strong lines, and is then fixed at this value for all the other lines. We also fix $n=2$ since these data are the signals at twice the modulation frequency. We see three lines in Fig.~\ref{fig:n2o_calib_line}. The two smaller ones have low optical depth and their profiles are similar to the second derivative of the Voigt profile. The larger one has high optical depth resulting in saturation of the absorption. This broadens the profile and introduces an extra dip at the centre of the profile which corresponds to the flat top of the saturated absorption line where the second derivative tends towards zero. This central dip becomes deeper and broader as the optical depth increases. Looking at many lines, we find that the fits give large variations for the individual Gaussian and Lorentzian widths, but a consistent value for their quadrature. The mean value is a full width at half maximum (FWHM) of 36~MHz. If we instead fix the FWHM of the Gaussian profile to 31.8~MHz, which is the expected value due to Doppler broadening, the typical Lorentzian width is 12~MHz. This is close to the expected FWHM due to pressure broadening alone, which is 10.5~MHz, where we have used a typical value given by HITRAN~\cite{Gordon2022} for the self-broadening coefficient of these lines, 3~GHz/atm. The high signal-to-noise ratio and the accurate modelling of the line profile allows us to find relative line centres with fitting uncertainties below $10^{-4}$~cm$^{-1}$ for almost all lines.

\subsection{Data reduction}

\begin{figure}[t]
    \centering
    \includegraphics[width=\linewidth]{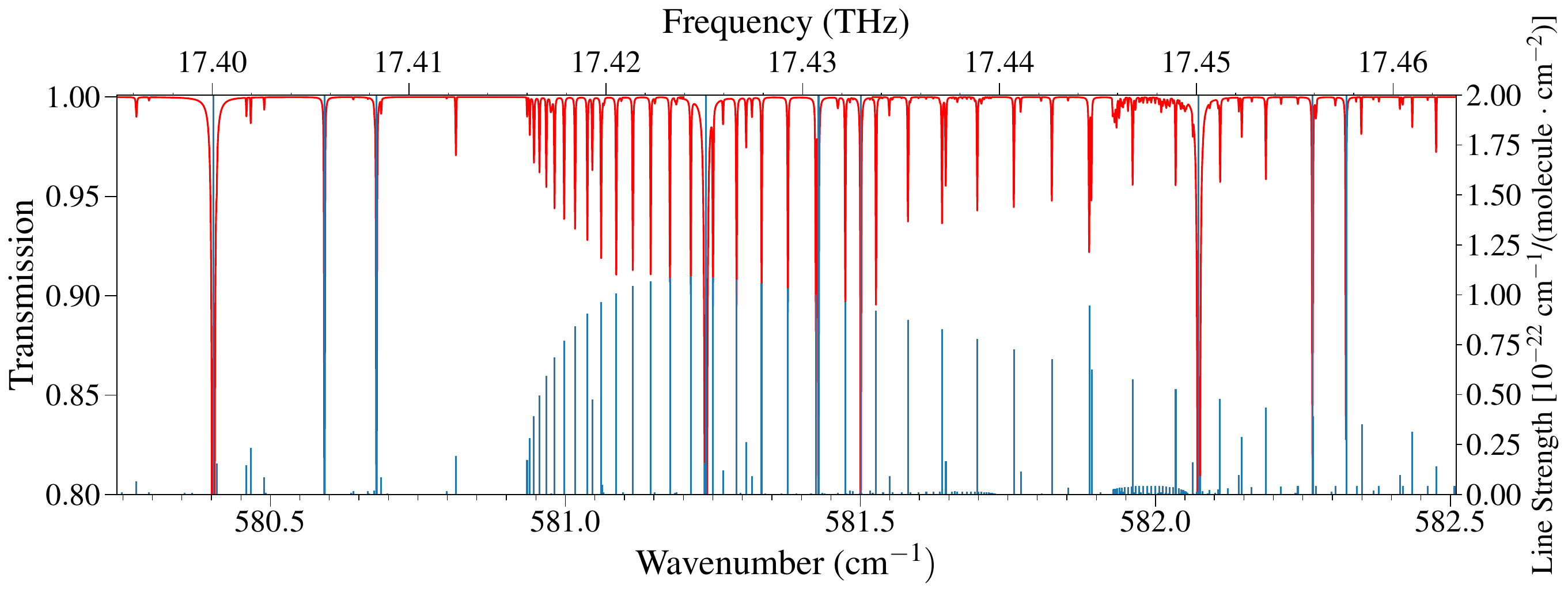}
    \caption{Comparison between our measured N$_2$O transitions and those in the HITRAN~\citep{Gordon2022} database. The red line is the fitted absorption spectrum converted from the original wavelength modulated spectrum. For this dataset, the left hand axis gives the fraction of incident power transmitted through the cell. The blue sticks are the N$_2$O transitions from HITRAN~\citep{Gordon2022} and are associated with the line strength given on the right axis. The strongest $\nu_2$ lines are much stronger than the other lines, having a height about 35 times the vertical scale of this figure.}
    \label{fig:fit_hitran}
\end{figure}

We have observed 165 N$_2$O lines in the frequency range from 580.27 to 582.48~cm$^{-1}$. They are listed in Tab.~\ref{tab:n2odata}. Figure \ref{fig:fit_hitran} shows the complete spectrum, where we have converted from the 2f modulation data back to the original absorption spectrum using the fitted Voigt parameters. This conversion is done by plotting $\sum_i e^{-\alpha_i(f-f_i)}$ where the sum is over all lines $i$, $f$ is the wavenumber, $f_i$ is the wavenumber of line $i$, and $\alpha_i$ is given by Eq.~(\ref{eq:alpha}) with $\sigma_{1,2}$ set to the values found from the fit to line $i$. Several strong lines, where the direct absorption is easily measured alongside the 2f data, are used to convert the amplitudes of the lines to the absolute transmission, thereby calibrating the vertical axis. We calibrate the frequency axis using the cavity data and three calibration lines from the NIST database~\citep{NIST118}, namely the P(8), P(9) and P(10) lines of the $(01^10)-(000)$ band of $^{14}$N$_2^{16}$O. The cavity data determines the frequency scale between each of the calibration lines. To treat the cavity data, we first remove the change in amplitude of the cavity fringes across the scans by normalizing the cavity data to the laser power. The power is determined from the baseline of the raw N$_2$O absorption spectrum. The resulting cavity fringes are almost sinusoidal because the cavity has such a low finesse. We apply a low-pass filter to the cavity data to remove higher order components, and then fit a sinusoidal model to segments of the data. These segments typically contain 2-7 cavity fringes and are short enough that non-linearity in the conversion from QCL current to frequency is negligible. After piecing the segments together, a cavity phase can be assigned to every point in the scan. The phase difference between two calibration lines, whose frequencies are known, provides the conversion from phase to frequency. In this way, the frequency of every line in the spectrum is determined. The full list of measured frequencies is given in the second column of Tab.~\ref{tab:n2odata}.

Next we consider the uncertainties in determining the centre frequencies of all the lines. As detailed in \ref{appendix:errorAnalysis}, there are four main sources of statistical uncertainty. The first comes from the calibration lines which each have a $1\sigma$ uncertainty of 1.5~MHz. The second comes from the fits to the spectral lines. This uncertainty varies enormously between the weak and strong lines. It is about 2~MHz for the weakest lines that we have measured, and about 100 times smaller than this for the strongest lines. These uncertainties are the ones returned by the fit algorithm so rely on the underlying model being sufficiently accurate. Where the lines are strong, we have compared frequency intervals determined from fits to the 2f modulation data to those found from fits to the direct absorption data. The fits to the 2f modulation data always return smaller uncertainties than fits to direct absorption data, as we would expect since the signal-to-noise ratio is much higher for the modulation data. Frequency intervals differ between the two methods by up to 0.5~MHz. Although this difference is small compared to other sources of uncertainty, it is statistically significant; typically the difference is 2-4 times the fit uncertainties. This suggests that the fit uncertainties may be under-estimated, most likely because the lineshape model fails to completely capture the exact lineshape of the 2f data. Indeed, this can be seen in figure \ref{fig:n2o_calib_line} where there is structure in the residuals. The probable under-estimate does not cause much concern since, for almost all lines, the fit errors are not the dominant source of uncertainty.

The third uncertainty comes from fitting to the cavity data and determining the cavity phase at each of the calibration lines and each of the lines of interest. This varies across the spectrum but is at most 0.4~MHz. The last source of uncertainty is due to the frequency drift of the cavity resulting from drifts of the laboratory temperature and pressure. We monitored these drifts by repeatedly scanning the same frequency region and measuring how the phase of the cavity changes with respect to an N$_2$O line over time. We also passed a 780~nm laser through the cavity and monitored how the cavity frequency changed relative to an atomic reference (a saturated absorption feature in Rb). The latter is more sensitive because of the shorter wavelength. In \ref{appendix:errorAnalysis}, we provide a detailed explanation of the effect of the cavity drift and provide some example data showing how the cavity length changed over time. During an hour-long period of monitoring, we measured a linear cavity drift of 30~kHz/s at 17~$\mu$m. The uncertainty in the spectral line frequencies due to such a linear drift is tiny because the calibration procedure almost perfectly eliminates the effect of the drift when both the drift and the frequency scan are linear in time. To produce a significant effect, the cavity length would need to change suddenly during the scan, or drift in one direction for part of the scan, and then the other direction for the rest. While monitoring the cavity, we did observe occasional sudden (few second) step-like changes in the cavity frequency with amplitudes of about 5~MHz at 17~$\mu$m. We suppose that such changes may have been occurring while acquiring the N$_2$O data and include the effect as an uncertainty. The uncertainty is largest for spectral lines that are far from a calibration line where they reach about 2.5~MHz.

Collisions induce both a broadening and a shift of the spectral lines. To estimate this pressure shift, we note that HITRAN gives shift coefficients in air ranging from $-15$ to $+3$~MHz/atm for all lines in this region, while Toth~\cite{Toth1993} gives an average self-shift coefficient of $-45$~MHz/atm for lines in the region 1800--2630~cm$^{-1}$. We did not find a value for the self-shift coefficients for our spectral region, but it should be similar to these values which give a negligible line shift for our operating pressure of roughly 350~Pa. We add all uncertainties in quadrature to obtain the total uncertainty on each line. These are given in parentheses in the second column of Tab.~\ref{tab:n2odata}, where the number indicates the $2\sigma$ uncertainty in the last digits. For most lines, this is between 3 and 5~MHz. The root mean square (rms) value of the set of uncertainties is 3.6~MHz. We note that, in most cases, the largest source of uncertainty comes from the calibration lines. This observation suggests that a more accurate absolute reference will be useful in improving  the accuracy of all the lines reported here to the sub-MHz level. 

The absolute pressure of N$_2$O was not calibrated in the present work, so we are not currently able to report spectral line intensities on an absolute scale. Instead, Tab.~\ref{tab:n2odata} (last column) gives the relative amplitude of each line in the spectrum which is the fitted value of the amplitude parameter of the Voigt function.

\subsection{Comparison with other data}

Figure \ref{fig:fit_hitran}, compares our measured spectrum to a model spectrum derived from the HITRAN database~\citep{Gordon2022}. The strengths of these lines span 4 orders of magnitude, and we have chosen a vertical scale where the weakest lines are just visible; the line strengths of the most intense lines are about 35 times larger than the scale of the graph. The measured spectrum is in good agreement with HITRAN for both the frequencies and line strengths. Table \ref{tab:n2odata} lists the frequencies, isotopologues and quantum numbers of all the lines measured in this work and compares our measured frequencies to the ones given in the HITRAN database and the NIST table of calibration lines. Our measurements agree with the values given in HITRAN to better than $6\times 10^{-4}$~cm$^{-1}$ (18~MHz) for 148 of the 165 lines, and the rms deviation between HITRAN and our measurements is $4.7\times 10^{-4}$~cm$^{-1}$ (14~MHz).

\subsection{Molecular parameters}

In the literature, three different approaches have been used to model the rotational structure of the vibrational bands of N$_2$O. They differ in the way they treat the splitting of states with $l \ne 0$ into e and f levels, which we call the $l$-doubling. In the first approach the e and f states are treated as separate vibrational states, each with their own term energy and set of rotational constants. This tends to introduce more parameters than are required. The second method, used for example in \cite{Olson1981}, describes the energy using the empirical formula
\begin{align}
    E_v &= G_v + B_v J(J+1) - D_v (J(J+1)-l^2)^2 + H_v (J(J+1)-l^2)^3 \nonumber\\
    &\pm \frac{1}{2}q_v J(J+1) \mp \frac{1}{2}q_{vJ} J^2(J+1)^2 \pm \frac{1}{2}q_{vJJ} J^3(J+1)^3.
    \label{eq:vibrational_energies}
\end{align}
Here, the subscript $v$ stands for the set of quantum numbers $\nu_1,\nu_2,l,\nu_3$, $G_v$ is the term energy, $B_v$, $D_v$ and $H_v$ are rotational constants and $q_v$, $q_{vJ}$ and $q_{vJJ}$ are $l$-doubling constants (set to zero when $l=0$). The third approach, used in \cite{Maki1992}, includes matrix elements of the $l$-doubling Hamiltonian that couples states differing in $l$ by 2 units. This is more consistent, but has the disadvantage that the energies of levels in one vibrational state depend on the parameters of a different vibrational state. We choose to use the second method and fit to our data using Eq.~(\ref{eq:vibrational_energies}).

We analyze those lines arising from the main isotopologue, $^{14}$N$_2^{16}$O. There is insufficient data for the other isotopologues. For most of the vibrational states in our analysis, the rotational and $l$-doubling parameters have been established to high precision through microwave spectroscopy. The parameters are summarized in Table 1 of \cite{Olson1981}. For the states $(00^00)$, $(01^10)$, $(02^00)$, $(02^20)$, $(03^10)$, $(03^30)$, $(11^10)$ and $(10^00)$, we fix $B_v$, $D_v$, $H_v$, $q_v$, $q_{vJ}$ and $q_{vJJ}$ to the values given in \cite{Olson1981}, leaving only the $G_v$ as free parameters. The rotational and $l$-doubling parameters for $(04^20)$ and $(12^00)$ are not provided in \cite{Olson1981} so we leave them all as free parameters in our fits and determine them from our own data. We fit each band separately to Eq.~(\ref{eq:vibrational_energies}). The output from these fits are the values of the band centre frequencies $\nu_0 = G_{v'} - G_{v}$, together with the rotational and $l$-doubling parameters for $(04^20)$ and $(12^00)$\footnote{$v'$ and $v$ are the vibrational quantum numbers of the upper and lower states}.

\begin{table}[tb]
    \centering
    \begin{tabular}{l|l|l|l}
        \hline
         Band & $v_0$ (cm$^{-1}$) & Lines & Inflation \\ \hline
         $(01^10) - (00^00)$ & 588.76773(8) & 3 & 1  \\

         $(02^00) - (01^10)$ & 579.36440(23) & 2 & 3.2  \\
         
         $(02^20) - (01^10)$ & 588.97656(9) & 6 & 1.7  \\
         
         $(03^10) - (02^00)$ & 580.93263(6) & 35 & 1.3  \\
        
         $(03^30) - (02^20)$ & 589.16739(13) & 6 & 2.5  \\

         $(03^10) - (02^20)$ & 571.32052(9) & 6 & 1.7  \\
       
         $(04^20) - (03^10)$ & 582.05610(7) & 51 & 1.3  \\
         
         $(11^10) - (10^00)$ & 595.36234(8) & 3 & 1  \\
        
         $(12^00) - (11^10)$ & 581.73044(10) & 21 & 1  \\
         \hline
    \end{tabular}
    \caption{Band centre frequencies, in cm$^{-1}$, for nine vibrational bands of $^{14}$N$_2^{16}$O. The number in the parenthesis is the 1$\sigma$ uncertainty in the last digits. The table also gives the number of lines used in the fit, and the inflation factor needed to bring the reduced chi-squared below 2.}
    \label{tab:spec_consts}
\end{table}

Table \ref{tab:spec_consts} lists the values of $\nu_0$ determined from the fits for nine vibrational bands of $^{14}$N$_2^{16}$O, along with their uncertainties. To judge the goodness of the fits, we use the reduced chi-squared parameter, $\chi_{\rm r}^2$, which is the sum of the squared standardized residuals divided by the number of degrees of freedom in the fit. For a good fit, we expect $\chi_{\rm r}^2 \approx 1$. For the two bands involving $(04^20)$ and $(12^00)$, where the rotational and $l$-doubling parameters are free parameters in the fit, we obtain values of $\chi_{\rm r}^2$ close to 1, indicating that equation (\ref{eq:vibrational_energies}) is an adequate model and that the frequency uncertainites on the individual lines are estimated well. However, for some other bands where we have relied on literature values of the rotational and $l$-doubling parameters, we found larger values of $\chi_{\rm r}^2$. This sometimes happens when combining molecular parameters from different sources, perhaps due to different analysis procedures or because different spectral regions are used to generate the parameters. When the model is not an adequate representation of the data, the uncertainties in the fit parameters will not be correct. To account for this, we inflate the uncertainties of the individual lines used in the fit until the value of $\chi_{\rm r}^2$ is reduced below 2. This inflation produces a corresponding increase in the uncertainties of the fit parameters. The inflation factor needed for each band is given in the last column of table \ref{tab:spec_consts}. In most, but not all, cases, only a minor inflation is needed. The uncertainties in the values of $\nu_0$ are the quadrature sums of the uncertainties in the calibration lines and the errors returned by the fitting routine. Values of $\nu_0$ for the first five bands in Tab.~\ref{tab:spec_consts} are also given in \cite{Olson1981}. Our values are consistent with \cite{Olson1981} and are up to 17 times more precise. 

From the $\nu_0$, we obtain the values of $G_v$ for each of the vibrational states. The only information needed beyond Tab.~\ref{tab:spec_consts} is the frequency of the $v_1$ mode, i.e. the value of $G_v$ for $(10^00)$. For this, we use the value given in Table 7 of \cite{Maki1992}\footnote{Reference \cite{Maki1992} has mistakenly swapped the $v_1$ and $v_3$ quantum numbers in Table 7 and in the associated spectral line assignments.}. Table \ref{tab:param_summary} summarizes the parameters of nine vibrational states of $^{14}$N$_2^{16}$O. Here, we present the values derived in the present work together with values given in Ref.~\cite{Olson1981}. In determining the uncertainties on the values of $G_v$ we have added together the uncertainties on the contributing $\nu_0$ in quadrature, assuming they are uncorrelated. The uncertainties on the parameters of $(04^20)$ and $(12^00)$ are those returned by the fitting routine. For $(03^10)$ we could determine two separate values of $G_v$, one from the analysis of the $(03^10)-(02^00)$ band and the other from the $(03^10)-(02^20)$ band. These two values were consistent within their errors, and we have taken their weighted mean. 

\begin{figure}[tb]
    \centering
    \includegraphics[width=\linewidth]{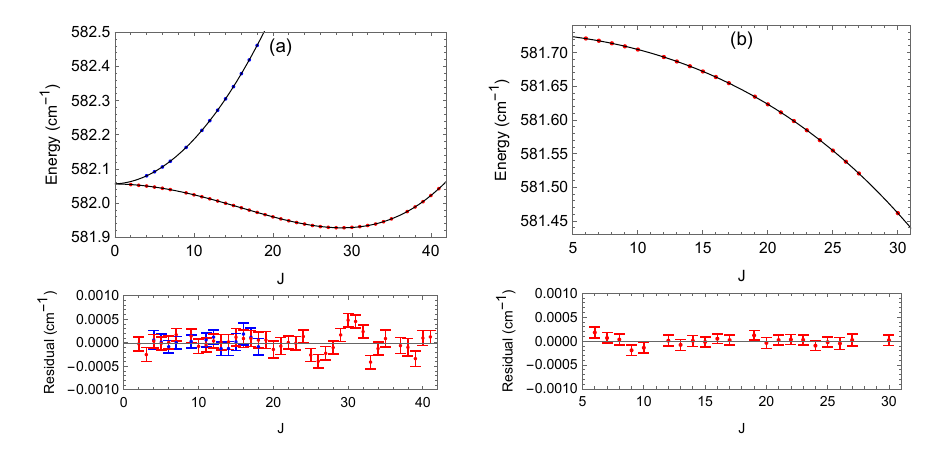}
    \caption{(a) Points: energies of lines in the Q-branch of the $(04^20)-(03^10)$ band; blue/red points have e/f symmetry in the lower level. Lines: fit to Eq.~(\ref{eq:vibrational_energies}). (b) Points: energies of lines in the Q-branch of $(12^00)-(11^10)$; all points have f symmetry in the lower level.  Line: fit to Eq.~(\ref{eq:vibrational_energies}). Residuals are shown beneath each plot.}
    \label{fig:fit_bands}
\end{figure}

Figure \ref{fig:fit_bands} shows the result of fitting Eq.~(\ref{eq:vibrational_energies}) to the Q branches of the $(04^20)-(03^10)$ and $(12^00)-(11^10)$ bands. These are the fits used to determine the parameters of $(04^20)$ and $(12^00)$ in Tab.~\ref{tab:param_summary}. The reduced chi-squared values for these fits are given in Tab.~\ref{tab:spec_consts}. The $(12^00)-(11^10)$ data fit very well, with residuals all consistent with zero and no structure evident in the residuals. For the $(04^20)-(03^10)$ band, the points with e symmetry in the lower level are in blue and fit very well, while the points with f symmetry in the lower level are in red and show a slight deviation from the model around the turning point of the parabola indicating a minor failure of the model in this region.

\begin{table}[tb]
    \centering
    \setlength\tabcolsep{4 pt}
    \footnotesize
    \begin{tabular}{c|c|c|c|c|c|c}
        \hline
         State & $G_v$ & $B_v$ & $10^7 D_v$  & $10^{14} H_v$  & $10^4 q_v$  & $10^9 q_{vJ}$  \\ \hline
         $(00^00)$ & 0 & 0.4190109955(58) & 1.76067(11) & -2.68(34) & -- & --   \\
         $(01^10)$ & 588.76773(8) & 0.419573601(11) & 1. 78884(l8) & -0.78(48) & 7.9200575(38) & 1 .0205(60)   \\
         $(02^00)$ & 1168.13213(24) & 0.419920179(20) & 2.48083(41) & 266.1(27) & -- & --   \\
         $(02^20)$ & 1177.74428(12) & 0.420124871(l5) & 1.51161(37) & -134.1(31) & 0.00374(31) & 61.047(73)   \\
         $(03^10)$ & 1749.06479(13) & 0.420331000(30) & 2.13767(81) & -19.0(73) & 14.95159(21) & 7.813(28)   \\
         $(03^30)$ & 1766.91168(17) & 0.420663959(82) & 1.5986(22) & -189(20) & -- & --   \\
         $(04^20)$ & 2331.12086(26) & 0.42076841(63) & 1.126(19) & -1380(23) & 0.0109(96) & 187(4)   \\
         $(11^10)$ & 1880.26563(14) & 0.417918418(16) & 1.73355(24) & 17.3(11) & 9.08316(19) & -3.005(23)  \\
         $(12^00)$ & 2461.99606(18) & 0.41814597(88) & 2.394(22) & 118(152) & -- & --  \\

         \hline
    \end{tabular}
    \caption{Molecular parameters of nine vibrational states of $^{14}$N$_2^{16}$O. The numbers in the parentheses are the 1$\sigma$ uncertainties in the last digits. All values are in cm$^{-1}$. The parameter $q_{vJJ}$ is negligible for all states except for $(02^20)$ where $q_{vJJ} = 2.755(60)\times 10^{-12}$~cm$^{-1}$. All values of $G_v$, and all parameters for $(04^20)$ and $(12^00)$ are from the present work. All other values are from \cite{Olson1981}.}
    \label{tab:param_summary}
\end{table}

\section{Conclusions}

Using a quantum cascade laser near 17~$\mu$m and wavelength modulation spectroscopy, we have determined the frequencies of 165 vibrational transitions of N$_2$O. We have presented a model for the lineshape of absorption features in wavelength modulation spectroscopy, and used this model to fit our spectral data. An optical cavity is used to calibrate the relative frequency scale, and a few calibration lines determined previously are used to fix the absolute frequency. This method transfers the precision of the calibration lines onto all other lines. Using our data alongside previous microwave measurements, we determine the term energies ($G_v$) of eight vibrational states of N$_2$O and the rotational and $l$-doubling parameters ($B_v$, $D_v$, $H_v$, $q_v$, $q_{vJ}$) of two states where this information was previously absent. These new spectroscopic data can be used to build more accurate line lists of N$_2$O in this spectral region, refining and extending existing databases. 

In the present work, we only measured the relative amplitudes of lines in the spectrum. The absolute spectral line intensities are important for simulating the transmission of light through atmospheres and could be measured in future work by calibrating the pressure of N$_2$O in the cell. 

The frequency uncertainty of the present measurements are dominated by the uncertainties of the calibration lines and drift in the optical cavity used to establish the frequency scale. In most cases, the determination of the line centre makes a negligible contribution to the overall uncertainty at the present level. In future work, this uncertainty could be reduced further by using sub-Doppler spectroscopy techniques. The cavity drift can be eliminated by stabilizing its temperature and pressure, and/or by using a frequency-stabilized laser in the visible to stabilise the cavity length. Ongoing work aims to improve calibration in this spectral region by building a frequency standard based on the fundamental vibrational transition in trapped, ultracold CaF molecules~\citep{Barontini2022}. Standards of this kind could be developed throughout the mid-infrared using the vibrational transitions of a wide range of ultracold molecules~\citep{Fitch2021b,Leung2022}. These standards can then be used to improve the precision of sets of spectral lines in common gases such as N$_2$O, using the methods described in the present work. For example, one QCL could be locked to a vibrational transition in the ultracold molecule, while a second QCL is used for spectroscopy of the type presented here, and their difference frequency measured using heterodyne techniques.

The data that support the findings of this study are openly available at the following URL/DOI: 10.5281/zenodo.14703486

\section*{Acknowledgements}
We are grateful to Roland Teissier and Alexei N. Baranov from IES, University of Montpellier, France, for supplying the quantum cascade laser. This work was supported by STFC and EPSRC under grants ST/T006234/1 and ST/W006197/1, and by Region Ile-de-France in the framework of DIM QuanTiP, and by the Imperial College London-CNRS 2021 PhD joint programme. Part of the work was carried out within the project 23FUN04 COMOMET, which is funded from the European Partnership on Metrology, co-financed from the European Union’s Horizon Europe Research and Innovation Programme and supported in the UK by UKRI (10130794).

\newcommand{\newblock}{}
\bibliographystyle{unsrt}
\bibliography{references}

\appendix

\section{Shape of wavelength modulated spectral lines}
\label{appendix:WMS}

In this appendix, we derive an expression for the lineshape of an absorption feature obtained from wavelength modulation spectroscopy. Our analysis extends that found in Refs. \citep{Arndt1965, Schilt2003}. It captures the effects of Doppler broadening and pressure broadening, lineshape changes due to the modulation of the wavelength, the effect of large optical depth, and the effect of simultaneous intensity and wavelength modulation.

Let $\omega$ be the angular frequency of the laser, $\omega_{\rm c}$ the centre frequency of the absorption feature, and $\gamma$ a quantity representing the linewidth of the absorption feature. The laser current is modulated at frequency $\Omega$ and this modulates both the laser intensity and the laser wavelength. We introduce a normalized detuning 
\begin{equation}
    \Delta = (\omega-\omega_{\rm c})/\gamma = \delta - \delta_{\rm m}\cos(\Omega t + \Phi).
\end{equation}
Here, $\delta$ is the detuning of the laser averaged over a modulation cycle and $\delta_{\rm m}$ is the modulation amplitude, both normalized to $\gamma$. The intensity of the light transmitted through the absorbing sample is $I_{\rm out} = I_{\rm in} \exp(-\alpha(\Delta))$ where $I_{\rm in}$ is the input intensity and $\alpha$ is the frequency-dependent optical depth of the sample. We assume that the input intensity varies linearly with the frequency, $I_{\rm in} = I_0(1+ p\delta - q \delta_{\rm m} \cos(\Omega t))$, where $I_0$ is the intensity on resonance, $p$ characterizes the intensity change due to a dc (or slowly-varying) offset, and $q$ characterizes the intensity change due to the modulation. $\Phi$ is a phase shift between the modulation of the wavelength and the intensity. With these definitions, the transmitted intensity is
\begin{equation}
    I_{\rm out}(\delta, t) = I_0\left(1+p\delta - q\delta_{\rm m}\cos(\Omega t)\right)\exp(-\alpha(\delta - \delta_{\rm m}\cos(\Omega t + \Phi))).
\end{equation}
We choose to re-write this in the form
\begin{equation}
    I_{\rm out}(\delta, t) = g_1(\delta - \delta_{\rm m}\cos(\Omega t + \Phi)) + g_2(\delta - \delta_{\rm m}\cos(\Omega t + \Phi)), 
\end{equation}
where
\begin{align}
    g_1(x) &= I_0 p x\exp(-\alpha(x))\\
    g_2(x) &= I_0 c_2 \exp(-\alpha(x))
\end{align}
and
\begin{equation}
    c_2 = 1-2p\delta_{\rm m}\sin(\Phi/2)\sin(\Omega t + \Phi/2) + (p-q) \delta_{\rm m}\cos(\Omega t).\label{app:eq:c2}
\end{equation}

Our aim is to find the component of $I_{\rm out}$ oscillating at frequency $n \Omega$, where $n$ is an integer, since this is the output of the lock-in amplifier. To do this, we first take the Fourier transform of $I_{\rm out}(\delta)$, which we denote $S(\tau)$:
\begin{equation}
    S(\tau,t) = [G_1(\tau)+G_2(\tau)]\exp(-i \delta_{\rm m} \tau \cos(\Omega t+\Phi)). \label{app:eq:S}
\end{equation}
Here, $G_1(\tau)$ and $G_2(\tau)$ are the Fourier transforms of $g_1(\delta)$ and $g_2(\delta)$ and we have used the shift theorem of Fourier transforms. Next, we use the Jacobi-Anger expansion which has the general form $e^{i z \cos\theta} = J_0(z) + 2 \sum_{n=1}^{\infty} i^n J_n(z) \cos(n\theta)$, where $J_n$ is the Bessel function of the first kind. Applying this to Eq.(\ref{app:eq:S}) and using $J_n(-z)=(-1)^n J_n(z)$ we get
\begin{equation}
    S(\tau, t) = [G_1(\tau)+G_2(\tau)]\sum_{n=0}^{\infty} \epsilon_n (-i)^n J_n(\delta_{\rm m} \tau)\cos[n(\Omega t + \Phi)],
\end{equation}
where $\epsilon_0=1$ and $\epsilon_n = 2$ for all $n>0$. We now take the inverse Fourier transform, noting that the cosine factor does not depend on $\tau$, to obtain
\begin{equation}
I_{\rm out}(\delta, t) = s_1(\delta,t) + s_2(\delta, t)
\end{equation}
where
\begin{equation}
    s_k(\delta,t) = \sum_{n=0}^{\infty} s_{k,n}(\delta) \cos[n(\Omega t + \Phi)],
\end{equation}
and
\begin{equation}
   s_{k,n}(\delta) = \epsilon_n (-i)^n \frac{1}{2\pi}  \int_{-\infty}^{\infty} e^{i \delta \tau} J_n(\delta_{\rm m} \tau) G_k(\tau) d\tau. 
   \label{app:eq:skn}
\end{equation}
The integral in Eq.~(\ref{app:eq:skn}) is the inverse Fourier transform of the product of $J_n$ and $G_k$. Applying the convolution theorem, this can be written as
\begin{equation}
    s_{k,n}(\delta) = \epsilon_n (-i)^n \frac{1}{\delta_{\rm m}}  \int_{-\infty}^{\infty} j_n\left(\frac{\delta'}{\delta_{\rm m}}\right) g_k(\delta-\delta') d\delta',
\end{equation}
where $j_n$ is the Fourier transform of the Bessel function, $j_n(x) = 2(-i)^n \cos(n\cos^{-1}(x))/\sqrt{1-x^2}$ for $|x|<1$ and zero otherwise. So we have
\begin{equation}
    s_{k,n}(\delta) = 2 \epsilon_n (-1)^n \int_{-\delta_{\rm m}}^{\delta_{\rm m}} \frac{\cos(n\cos^{-1}(\delta'/\delta_{\rm m}))}{\sqrt{\delta_{\rm m}^2 - \delta'^2}}g_k(\delta-\delta') d\delta'.
\end{equation}

The decomposition into a sum of harmonics of $\Omega$ is complicated by the presence of trigonometric functions of $\Omega t$ in the expression for $c_2$, Eq.~(\ref{app:eq:c2}). This can be sorted out using
\begin{align}
    &c_2 \cos[n(\Omega T + \Phi)] =\cos[n(\Omega T + \Phi)] \nonumber\\&+ \frac{1}{2}\delta_m p \left( \cos[(n-1)(\Omega T + \Phi)] + \cos[(n+1)(\Omega T + \Phi)] \right) \nonumber \\&- \frac{1}{2}\delta_m q \left( \cos[(n-1)\Omega T + n\Phi] + \cos[(n+1)\Omega T + n\Phi] \right).
\end{align}
This shows that the term oscillating at $n\Omega$ will have contributions from $s_{2,n\pm 1}$ as well as from $s_{2,n}$.

These equations can be used to determine the lineshape in the most general cases. In the case where the intensity and frequency oscillate in phase ($\Phi=0$) and the intensity changes in the same way for slow and fast changes ($p=q$), the result simplifies because $c_2 = 1$. In this case, the component of the output signal oscillating at frequency $n\Omega$ is
\begin{equation}
    \frac{I_{{\rm out},n}}{I_0} = 2\epsilon_n (-1)^n \int_{-\delta_{\rm m}}^{\delta_{\rm m}} \frac{\cos\left(n\cos^{-1}\left(\frac{\delta'}{\delta_{\rm m}}\right)\right)}{\sqrt{\delta_{\rm m}^2 - \delta'^2}}(1+p\delta-p\delta') e^{-\alpha(\delta-\delta')} d\delta'.
    \label{eq:demodulatedSignal}
\end{equation}

The function $\alpha(\delta)$ should be chosen according to the dominant broadening mechanisms. In this paper, we have used the Voigt function. We found Eq.~(\ref{eq:demodulatedSignal}) to be an adequate representation of the lineshapes we measure across the full range of optical depths. We note that others have used expressions that avoid the two assumptions ($\Phi=0$, $p=q$) used above~\cite{Schilt2003, Tran2024}.

\begin{figure}[t]
    \centering
    \includegraphics[width=\linewidth]{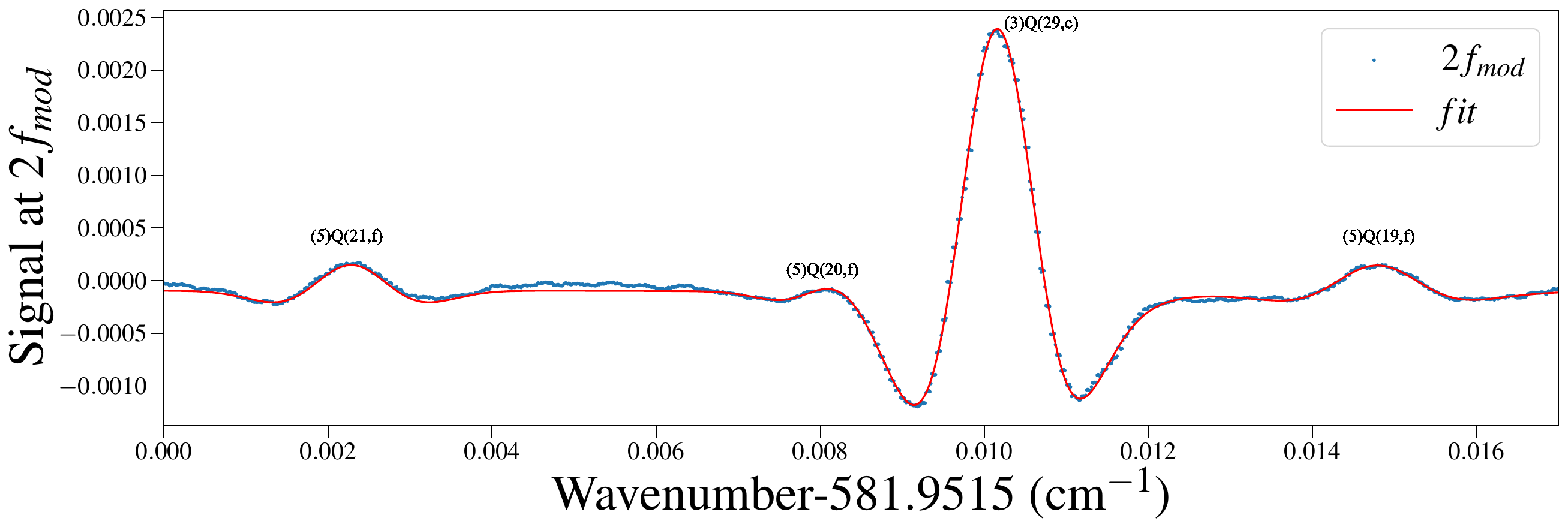}
    \caption{Example of a fit to $2f_{\rm mod}$ data with a partially overlapped line. The labelling is the same as in Fig.~\ref{fig:n2o_spectra_freq}.}
    \label{fig:fit_demo}
\end{figure}

Figure \ref{fig:n2o_calib_line} shows one example of fitting our model to the $2f_{\rm mod}$ data in a case where the lines are relatively strong. Figure \ref{fig:fit_demo} shows another example where the lines are 2-4 orders of magnitude weaker and two of the lines are partly overlapped. Specifically, the very weak line labelled $(5)Q(20,f)$ appears in the low frequency wing of the much stronger line labelled $(3)Q(29,e)$, which is itself about 100 times weaker than the strong line seen in Fig.~\ref{fig:n2o_calib_line}. Nevertheless, the model still fits well to the data and finds the centre frequencies of all 4 lines with relatively small uncertainty.

\section{Frequency uncertainties}
\label{appendix:errorAnalysis}

We record the N$_2$O spectrum and the cavity transmission simultaneously, giving two sets of data with a common abscissa. The abscissa may be the QCL current,  or it may be time since the current is scanned linearly - we will call it $t$. The laser frequency is a function of $t$ but may not be a perfectly linear function. We fit to the cavity transmission data to get the cavity phase as a function of $t$, which is proportional to the laser frequency, $\phi(t) = \alpha f(t) + \theta$. We use three calibration lines, but for any line in the spectrum we only use the two closest calibration lines. They have frequencies $f_i$ and corresponding cavity phases $\phi_i$, with $\phi_i = \alpha f_i + \theta$. With these definitions, the frequency can be expressed as
\begin{equation}
    f(t) =  \frac{f_1(\phi_2-\phi(t)) - f_2(\phi_1-\phi(t))}{\phi_2 - \phi_1}.
    \label{app:eq:fx}
\end{equation}
From Eq.~(\ref{app:eq:fx}) we find the frequency uncertainty to be
\begin{equation}
    \sigma_f = \sqrt{\sum_{i=1}^2\left(\frac{\phi-\phi_i}{\phi_2-\phi_1}\right)^2 (\sigma_{f_i}^2 + \beta^2 \sigma_{\phi_i}^2) + \beta^2\sigma_{\phi}^2 + \beta^2\left(\frac{d\phi}{d t}\right)^2 \sigma_t^2 },
\end{equation}
where $\beta = \frac{f_2-f_1}{\phi_2 - \phi_1}$. Here, $\sigma_{f_1}$ and $\sigma_{f_2}$ are the frequency uncertainties in the two calibration lines, $\sigma_{\phi_1}$ and $\sigma_{\phi_2}$ are the uncertainties in the cavity phases at the two calibration lines, $\sigma_\phi$ is the uncertainty in the cavity phase at the spectral line being measured, and $\sigma_t$ is the uncertainty in determining the line centre of the spectral line on the $t$-axis.

\begin{figure}[t]
    \centering
    \includegraphics[width=\linewidth]{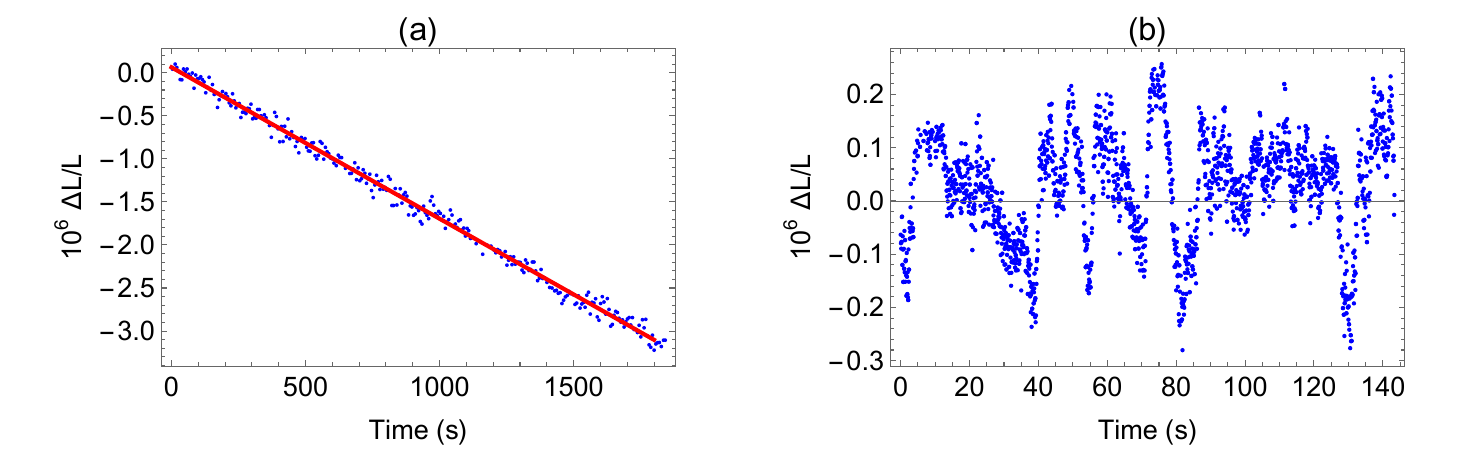}
    \caption{Fractional change in the cavity length as a function of time shown on (a) long and (b) short timescales.}
    \label{fig:cavity_drift}
\end{figure}

In addition, we need to account for the uncertainty from the drift of the cavity's optical path length, which arises due to changes in temperature and pressure. Let us first suppose that the cavity length drifts linearly in time at a rate $\gamma$, and the frequency is scanned linearly in time. Recognizing that the cavity phase is proportional to the frequency of the light multiplied by the length of the cavity, we find that Eq.~(\ref{app:eq:fx}) is modified to
\begin{equation}
    f(t) =  \frac{f_1(t_2-t)(L+\gamma t_1) - f_2(t_1-t)(L+\gamma t_2)}{(t_2 - t_1)(L+\gamma t)},
    \label{app:eq:fx2}
\end{equation}
where $L$ is the nominal length of the cavity and $t_{1,2}$ are the times in the scan where the calibration lines appear. We treat the cavity drift as an uncertainty by assuming that its mean value is zero but its variance is not. This leads to a frequency uncertainty due to cavity drift of
\begin{equation}
    \sigma_f = \frac{(f_2-f_1)}{(t_2 - t_1)}(t-t_1)(t_2-t) \frac{1}{L} \sigma_{\gamma}.
\end{equation}

We monitored the change in length of the cavity over time by passing a 780 nm laser through the cavity and measuring how the cavity frequency changed relative to a saturated absorption feature in Rb. Figure \ref{fig:cavity_drift}(a) shows  the result on a 30 minute timescale, where we see a linear drift of the cavity length. This drift corresponds to a frequency drift of the cavity at 17~$\mu$m of 30~kHz/s. This translates to $\sigma_f < 2$~kHz, which is negligible. The uncertainty is tiny because the calibration procedure almost eliminates the effect of the drift when both the drift and the frequency scan are linear in time. The scan is not quite linear, but the non-linearity is small enough that the uncertainty remains negligible.

To produce a significant effect, the cavity has to drift in a different way. For example, suppose that in scanning from $f_1$ to $f_2$, the cavity length expands linearly during the first half of the scan and then contracts linearly during the second half so as to return to its original length by the end of the scan. This conspiratorial behaviour produces a much larger effect. The uncertainty is
\begin{equation}
    \sigma_f = \tau|\sigma_{\frac{df}{dt}}|
    \label{eq:cavity_jump}
\end{equation}
where $\frac{df}{dt}$ is the drift of the cavity resonance frequency and $\tau$ is the time to/from the nearest calibration line, i.e. $(t-t_1)$ or $(t_2-t)$, whichever is smaller. A sudden step change in the length of the cavity part way through a scan has a similar effect. Figure \ref{fig:cavity_drift}(b) is an example of the length changes over 140~s. We do see some sudden changes, corresponding to frequency jumps of about 5~MHz at 17~$\mu$m. We use this information, together with Eq.~(\ref{eq:cavity_jump}), to estimate the uncertainty in the frequency of the N$_2$O lines due to cavity instability.

\newgeometry{margin=1cm} 
\begin{landscape}
\centering
\scriptsize

\section{Table of frequencies}
\label{appendix:frequencyTable}

\begin{longtable}[!ht]{|l|l|l|l|l|l|l|l|l|l|l|l|l|l|l|l|l|l|l|l|l|}
    \hline
        iso & wn (observed) & wn (ht) & wn (nist) & o-ht & o-nist & f(observed) & v1' & v2' & l' & v3' & v1 & v2 & l & v3 & PQR & J' & J & p & ov & amp \\ \hline
        44 & 580.272780(098) & 580.27260 & ~ & 0.000180 & ~ & 17.3961403 & 0 & 1 & 1 & 0 & 0 & 0 & 0 & 0 & P & 4 & 5 & e & s & 0.0039 \\ 
        41 & 580.294380(101) & 580.29432 & ~ & 0.000060 & ~ & 17.3967879 & 0 & 4 & 0 & 0 & 0 & 3 & 1 & 0 & R & 8 & 7 & e & s & 0.0004 \\ 
        41 & 580.403216(098) & 580.40323 & 580.403216 & -0.000015 & 0.000000 & 17.4000507 & 0 & 1 & 1 & 0 & 0 & 0 & 0 & 0 & P & 9 & 10 & e & s & 1.6335 \\ 
        41 & 580.409019(099) & 580.40913 & 580.409125 & -0.000111 & -0.000106 & 17.4002246 & 1 & 1 & 1 & 0 & 1 & 0 & 0 & 0 & P & 17 & 18 & e & s & 0.0051 \\ 
        43 & 580.459383(102) & 580.45934 & ~ & 0.000043 & ~ & 17.4017345 & 0 & 1 & 1 & 0 & 0 & 0 & 0 & 0 & P & 5 & 6 & e & s & 0.0019 \\ 
        42 & 580.466629(102) & 580.46661 & ~ & 0.000019 & ~ & 17.4019518 & 0 & 1 & 1 & 0 & 0 & 0 & 0 & 0 & R & 6 & 5 & e & s & 0.0024 \\ 
        41 & 580.489628(116) & 580.48964 & ~ & -0.000012 & ~ & 17.4026412 & 0 & 3 & 1 & 0 & 0 & 2 & 2 & 0 & R & 11 & 10 & e & s & 0.0012 \\ 
        41 & 580.591669(134) & 580.59218 & ~ & -0.000511 & ~ & 17.4057004 & 0 & 2 & 2 & 0 & 0 & 1 & 1 & 0 & P & 9 & 10 & f & s & 0.0824 \\ 
        42 & 580.640513(130) & 580.64099 & ~ & -0.000477 & ~ & 17.4071647 & 0 & 2 & 2 & 0 & 0 & 1 & 1 & 0 & R & 6 & 5 & f & s & 0.0003 \\ 
        42 & 580.665053(129) & 580.66533 & ~ & -0.000277 & ~ & 17.4079003 & 0 & 2 & 2 & 0 & 0 & 1 & 1 & 0 & R & 6 & 5 & e & s & 0.0001 \\ 
        45 & 580.676511(137) & 580.67613 & ~ & 0.000381 & ~ & 17.4082439 & 0 & 1 & 1 & 0 & 0 & 0 & 0 & 0 & P & 6 & 7 & e & s & 0.0006 \\ 
        41 & 580.679905(120) & 580.67969 & ~ & 0.000215 & ~ & 17.4083456 & 0 & 2 & 2 & 0 & 0 & 1 & 1 & 0 & P & 9 & 10 & e & s & 0.0820 \\ 
        41 & 580.687841(120) & 580.68743 & ~ & 0.000411 & ~ & 17.4085835 & 0 & 3 & 1 & 0 & 0 & 2 & 2 & 0 & R & 11 & 10 & f & s & 0.0015 \\ 
        42 & 580.798609(122) & 580.79890 & ~ & -0.000291 & ~ & 17.4119043 & 0 & 2 & 0 & 0 & 0 & 1 & 1 & 0 & R & 14 & 13 & e & s & 0.0001 \\ 
        41 & 580.814320(111) & 580.81411 & ~ & 0.000210 & ~ & 17.4123753 & 0 & 3 & 3 & 0 & 0 & 2 & 2 & 0 & P & 9 & 10 & e & o & 0.0045 \\ 
        41 & 580.814320(111) & 580.81494 & ~ & -0.000620 & ~ & 17.4123753 & 0 & 3 & 3 & 0 & 0 & 2 & 2 & 0 & P & 9 & 10 & f & o & 0.0045 \\ 
        41 & 580.935105(106) & 580.93517 & ~ & -0.000065 & ~ & 17.4159963 & 0 & 3 & 1 & 0 & 0 & 2 & 0 & 0 & Q & 1 & 1 & e & s & 0.0018 \\ 
        41 & 580.939720(105) & 580.93980 & ~ & -0.000080 & ~ & 17.4161347 & 0 & 3 & 1 & 0 & 0 & 2 & 0 & 0 & Q & 2 & 2 & e & s & 0.0046 \\ 
        41 & 580.946644(105) & 580.94675 & ~ & -0.000106 & ~ & 17.4163422 & 0 & 3 & 1 & 0 & 0 & 2 & 0 & 0 & Q & 3 & 3 & e & s & 0.0103 \\ 
        41 & 580.955991(105) & 580.95602 & ~ & -0.000029 & ~ & 17.4166225 & 0 & 3 & 1 & 0 & 0 & 2 & 0 & 0 & Q & 4 & 4 & e & s & 0.0103 \\ 
        41 & 580.967635(103) & 580.96762 & ~ & 0.000015 & ~ & 17.4169715 & 0 & 3 & 1 & 0 & 0 & 2 & 0 & 0 & Q & 5 & 5 & e & s & 0.0119 \\ 
        41 & 580.975060(110) & 580.97595 & ~ & -0.000890 & ~ & 17.4171941 & 0 & 4 & 4 & 0 & 0 & 3 & 3 & 0 & P & 9 & 10 & f & s & 0.0030 \\ 
        41 & 580.977014(119) & 580.97621 & ~ & 0.000804 & ~ & 17.4172527 & 0 & 4 & 4 & 0 & 0 & 3 & 3 & 0 & P & 9 & 10 & e & s & 0.0019 \\ 
        41 & 580.981594(102) & 580.98155 & ~ & 0.000044 & ~ & 17.4173900 & 0 & 3 & 1 & 0 & 0 & 2 & 0 & 0 & Q & 6 & 6 & e & s & 0.0170 \\ 
        41 & 580.997890(103) & 580.99780 & ~ & 0.000090 & ~ & 17.4178786 & 0 & 3 & 1 & 0 & 0 & 2 & 0 & 0 & Q & 7 & 7 & e & s & 0.0180 \\ 
        41 & 581.016500(102) & 581.01640 & ~ & 0.000100 & ~ & 17.4184365 & 0 & 3 & 1 & 0 & 0 & 2 & 0 & 0 & Q & 8 & 8 & e & s & 0.0187 \\ 
        41 & 581.037390(101) & 581.03733 & ~ & 0.000060 & ~ & 17.4190627 & 0 & 3 & 1 & 0 & 0 & 2 & 0 & 0 & Q & 9 & 9 & e & s & 0.0202 \\ 
        41 & 581.045788(101) & 581.04559 & 581.045815 & 0.000198 & -0.000027 & 17.4193145 & 0 & 2 & 0 & 0 & 0 & 1 & 1 & 0 & R & 2 & 1 & e & s & 0.0103 \\ 
        41 & 581.060616(115) & 581.06062 & ~ & -0.000004 & ~ & 17.4197590 & 0 & 3 & 1 & 0 & 0 & 2 & 0 & 0 & Q & 10 & 10 & e & s & 0.0240 \\ 
        44 & 581.062542(117) & 581.06226 & ~ & 0.000282 & ~ & 17.4198168 & 0 & 1 & 1 & 0 & 0 & 0 & 0 & 0 & P & 3 & 4 & e & s & 0.0006 \\ 
        41 & 581.065240(131) & 581.06450 & ~ & 0.000740 & ~ & 17.4198977 & 0 & 4 & 2 & 0 & 0 & 3 & 3 & 0 & R & 20 & 19 & f & s & 0.0016 \\ 
        41 & 581.086009(105) & 581.08626 & ~ & -0.000251 & ~ & 17.4205203 & 0 & 3 & 1 & 0 & 0 & 2 & 0 & 0 & Q & 11 & 11 & e & s & 0.0276 \\ 
        41 & 581.094567(114) & 581.09753 & ~ & -0.002963 & ~ & 17.4207769 & 0 & 4 & 2 & 0 & 0 & 3 & 3 & 0 & R & 20 & 19 & e & s & 0.0005 \\ 
        41 & 581.114122(102) & 581.11427 & ~ & -0.000148 & ~ & 17.4213631 & 0 & 3 & 1 & 0 & 0 & 2 & 0 & 0 & Q & 12 & 12 & e & s & 0.0257 \\ 
        41 & 581.144577(101) & 581.14465 & ~ & -0.000073 & ~ & 17.4222761 & 0 & 3 & 1 & 0 & 0 & 2 & 0 & 0 & Q & 13 & 13 & e & s & 0.0254 \\ 
        41 & 581.151653(105) & 581.15140 & ~ & 0.000253 & ~ & 17.4224883 & 0 & 4 & 0 & 0 & 0 & 3 & 1 & 0 & R & 9 & 8 & e & s & 0.0011 \\ 
        41 & 581.177273(103) & 581.17741 & ~ & -0.000137 & ~ & 17.4232563 & 0 & 3 & 1 & 0 & 0 & 2 & 0 & 0 & Q & 14 & 14 & e & s & 0.0270 \\ 
        41 & 581.185699(117) & 581.18565 & ~ & 0.000049 & ~ & 17.4235089 & 1 & 2 & 2 & 0 & 1 & 1 & 1 & 0 & P & 15 & 16 & f & s & 0.0007 \\ 
        43 & 581.187960(119) & 581.18819 & ~ & -0.000230 & ~ & 17.4235767 & 0 & 2 & 0 & 0 & 0 & 1 & 1 & 0 & R & 8 & 7 & e & s & 0.0020 \\ 
        41 & 581.212526(099) & 581.21257 & ~ & -0.000044 & ~ & 17.4243132 & 0 & 3 & 1 & 0 & 0 & 2 & 0 & 0 & Q & 15 & 15 & e & s & 0.0278 \\ 
        41 & 581.236040(128) & 581.23593 & 581.235921 & 0.000110 & 0.000119 & 17.4250183 & 1 & 1 & 1 & 0 & 1 & 0 & 0 & 0 & P & 16 & 17 & e & s & 0.0094 \\ 
        41 & 581.238050(109) & 581.23806 & 581.238050 & -0.000014 & 0.000000 & 17.4250784 & 0 & 1 & 1 & 0 & 0 & 0 & 0 & 0 & P & 8 & 9 & e & s & 1.4866 \\ 
        41 & 581.250108(110) & 581.25012 & ~ & -0.000012 & ~ & 17.4254399 & 0 & 3 & 1 & 0 & 0 & 2 & 0 & 0 & Q & 16 & 16 & e & s & 0.0274 \\ 
        43 & 581.267218(096) & 581.26721 & ~ & 0.000008 & ~ & 17.4259528 & 0 & 1 & 1 & 0 & 0 & 0 & 0 & 0 & P & 4 & 5 & e & s & 0.0031 \\ 
        41 & 581.290115(096) & 581.29009 & ~ & 0.000025 & ~ & 17.4266392 & 0 & 3 & 1 & 0 & 0 & 2 & 0 & 0 & Q & 17 & 17 & e & s & 0.0300 \\ 
        42 & 581.306275(097) & 581.30626 & ~ & 0.000015 & ~ & 17.4271237 & 0 & 1 & 1 & 0 & 0 & 0 & 0 & 0 & R & 7 & 6 & e & s & 0.0071 \\ 
        41 & 581.316089(097) & 581.31608 & ~ & 0.000009 & ~ & 17.4274179 & 0 & 3 & 1 & 0 & 0 & 2 & 2 & 0 & R & 12 & 11 & e & s & 0.0030 \\ 
        41 & 581.332543(097) & 581.33247 & ~ & 0.000073 & ~ & 17.4279112 & 0 & 3 & 1 & 0 & 0 & 2 & 0 & 0 & Q & 18 & 18 & e & s & 0.0282 \\ 
        41 & 581.377275(099) & 581.37730 & ~ & -0.000025 & ~ & 17.4292522 & 0 & 3 & 1 & 0 & 0 & 2 & 0 & 0 & Q & 19 & 19 & e & s & 0.0311 \\ 
        41 & 581.424507(100) & 581.42457 & ~ & -0.000063 & ~ & 17.4306682 & 0 & 3 & 1 & 0 & 0 & 2 & 0 & 0 & Q & 20 & 20 & e & s & 0.0360 \\ 
        41 & 581.428922(100) & 581.42911 & ~ & -0.000188 & ~ & 17.4308006 & 0 & 2 & 2 & 0 & 0 & 1 & 1 & 0 & P & 8 & 9 & f & s & 0.0855 \\ 
        41 & 581.461740(107) & 581.46176 & ~ & -0.000020 & ~ & 17.4317844 & 1 & 2 & 0 & 0 & 1 & 1 & 1 & 0 & Q & 30 & 30 & f & s & 0.0026 \\ 
        41 & 581.474371(103) & 581.47429 & ~ & 0.000081 & ~ & 17.4321631 & 0 & 3 & 1 & 0 & 0 & 2 & 0 & 0 & Q & 21 & 21 & e & s & 0.0285 \\ 
        42 & 581.482067(110) & 581.48236 & ~ & -0.000293 & ~ & 17.4323938 & 0 & 2 & 2 & 0 & 0 & 1 & 1 & 0 & R & 7 & 6 & f & s & 0.0006 \\ 
        45 & 581.487453(118) & 581.48729 & ~ & 0.000163 & ~ & 17.4325553 & 0 & 1 & 1 & 0 & 0 & 0 & 0 & 0 & P & 5 & 6 & e & s & 0.0001 \\ 
        41 & 581.500515(106) & 581.50063 & ~ & -0.000115 & ~ & 17.4329469 & 0 & 2 & 2 & 0 & 0 & 1 & 1 & 0 & P & 8 & 9 & e & s & 0.0895 \\ 
        42 & 581.515805(120) & 581.51650 & 581.515805 & -0.000695 & ~ & 17.4334053 & 0 & 2 & 2 & 0 & 0 & 1 & 1 & 0 & R & 7 & 6 & e & s & 0.0001 \\ 
        41 & 581.520688(122) & 581.52091 & ~ & -0.000222 & ~ & 17.4335516 & 1 & 2 & 0 & 0 & 1 & 1 & 1 & 0 & Q & 27 & 27 & f & s & 0.0001 \\ 
        41 & 581.526320(106) & 581.52649 & ~ & -0.000170 & ~ & 17.4337205 & 0 & 3 & 1 & 0 & 0 & 2 & 0 & 0 & Q & 22 & 22 & e & s & 0.0248 \\ 
        41 & 581.538106(126) & 581.53846 & ~ & -0.000354 & ~ & 17.4340738 & 1 & 2 & 0 & 0 & 1 & 1 & 1 & 0 & Q & 26 & 26 & f & s & 0.0002 \\ 
        41 & 581.549688(103) & 581.55002 & ~ & -0.000332 & ~ & 17.4344210 & 0 & 3 & 1 & 0 & 0 & 2 & 2 & 0 & R & 12 & 11 & f & s & 0.0030 \\ 
        41 & 581.554614(104) & 581.55499 & ~ & -0.000376 & ~ & 17.4345687 & 1 & 2 & 0 & 0 & 1 & 1 & 1 & 0 & Q & 25 & 25 & f & s & 0.0003 \\ 
        41 & 581.570054(104) & 581.57055 & ~ & -0.000496 & ~ & 17.4350316 & 1 & 2 & 0 & 0 & 1 & 1 & 1 & 0 & Q & 24 & 24 & f & s & 0.0001 \\ 
        41 & 581.580833(102) & 581.58116 & ~ & -0.000327 & ~ & 17.4353547 & 0 & 3 & 1 & 0 & 0 & 2 & 0 & 0 & Q & 23 & 23 & e & s & 0.0173 \\ 
        41 & 581.584758(104) & 581.58517 & ~ & -0.000412 & ~ & 17.4354724 & 1 & 2 & 0 & 0 & 1 & 1 & 1 & 0 & Q & 23 & 23 & f & s & 0.0005 \\ 
        41 & 581.598454(104) & 581.59890 & ~ & -0.000446 & ~ & 17.4358830 & 1 & 2 & 0 & 0 & 1 & 1 & 1 & 0 & Q & 22 & 22 & f & s & 0.0001 \\ 
        41 & 581.611277(105) & 581.61176 & ~ & -0.000483 & ~ & 17.4362674 & 1 & 2 & 0 & 0 & 1 & 1 & 1 & 0 & Q & 21 & 21 & f & s & 0.0002 \\ 
        41 & 581.623229(114) & 581.62379 & ~ & -0.000561 & ~ & 17.4366258 & 1 & 2 & 0 & 0 & 1 & 1 & 1 & 0 & Q & 20 & 20 & f & s & 0.0001 \\ 
        41 & 581.634609(101) & 581.63502 & ~ & -0.000411 & ~ & 17.4369669 & 1 & 2 & 0 & 0 & 1 & 1 & 1 & 0 & Q & 19 & 19 & f & s & 0.0004 \\ 
        41 & 581.638073(101) & 581.63833 & ~ & -0.000257 & ~ & 17.4370707 & 0 & 3 & 1 & 0 & 0 & 2 & 0 & 0 & Q & 24 & 24 & e & s & 0.0187 \\ 
        41 & 581.644333(101) & 581.64460 & ~ & -0.000267 & ~ & 17.4372584 & 0 & 3 & 3 & 0 & 0 & 2 & 2 & 0 & P & 8 & 9 & e & o & 0.0071 \\ 
        41 & 581.644333(101) & 581.64520 & ~ & -0.000867 & ~ & 17.4372584 & 0 & 3 & 3 & 0 & 0 & 2 & 2 & 0 & P & 8 & 9 & f & o & 0.0071 \\ 
        41 & 581.654709(104) & 581.65522 & ~ & -0.000511 & ~ & 17.4375695 & 1 & 2 & 0 & 0 & 1 & 1 & 1 & 0 & Q & 17 & 17 & f & s & 0.0001 \\ 
        42 & 581.659056(103) & 581.65965 & ~ & -0.000594 & ~ & 17.4376998 & 0 & 2 & 0 & 0 & 0 & 1 & 1 & 0 & R & 15 & 14 & e & s & 0.0001 \\ 
        41 & 581.663774(102) & 581.66426 & ~ & -0.000486 & ~ & 17.4378413 & 1 & 2 & 0 & 0 & 1 & 1 & 1 & 0 & Q & 16 & 16 & f & s & 0.0010 \\ 
        41 & 581.672070(104) & 581.67261 & ~ & -0.000540 & ~ & 17.4380899 & 1 & 2 & 0 & 0 & 1 & 1 & 1 & 0 & Q & 15 & 15 & f & s & 0.0001 \\ 
        41 & 581.679809(115) & 581.68031 & ~ & -0.000501 & ~ & 17.4383220 & 1 & 2 & 0 & 0 & 1 & 1 & 1 & 0 & Q & 14 & 14 & f & s & 0.0002 \\ 
        41 & 581.686818(115) & 581.68739 & ~ & -0.000572 & ~ & 17.4385321 & 1 & 2 & 0 & 0 & 1 & 1 & 1 & 0 & Q & 13 & 13 & f & s & 0.0002 \\ 
        41 & 581.693398(115) & 581.69386 & ~ & -0.000462 & ~ & 17.4387294 & 1 & 2 & 0 & 0 & 1 & 1 & 1 & 0 & Q & 12 & 12 & f & s & 0.0005 \\ 
        41 & 581.697576(115) & 581.69801 & ~ & -0.000434 & ~ & 17.4388546 & 0 & 3 & 1 & 0 & 0 & 2 & 0 & 0 & Q & 25 & 25 & e & s & 0.0158 \\ 
        41 & 581.704514(107) & 581.70508 & ~ & -0.000566 & ~ & 17.4390626 & 1 & 2 & 0 & 0 & 1 & 1 & 1 & 0 & Q & 10 & 10 & f & s & 0.0015 \\ 
        41 & 581.709271(108) & 581.70986 & ~ & -0.000589 & ~ & 17.4392052 & 1 & 2 & 0 & 0 & 1 & 1 & 1 & 0 & Q & 9 & 9 & f & s & 0.0002 \\ 
        41 & 581.713778(111) & 581.71412 & ~ & -0.000342 & ~ & 17.4393403 & 1 & 2 & 0 & 0 & 1 & 1 & 1 & 0 & Q & 8 & 8 & f & s & 0.0001 \\ 
        41 & 581.717580(110) & 581.71787 & ~ & -0.000290 & ~ & 17.4394543 & 1 & 2 & 0 & 0 & 1 & 1 & 1 & 0 & Q & 7 & 7 & f & s & 0.0001 \\ 
        41 & 581.720964(113) & 581.72112 & ~ & -0.000156 & ~ & 17.4395558 & 1 & 2 & 0 & 0 & 1 & 1 & 1 & 0 & Q & 6 & 6 & f & s & 0.0001 \\ 
        41 & 581.759976(097) & 581.76021 & ~ & -0.000234 & ~ & 17.4407253 & 0 & 3 & 1 & 0 & 0 & 2 & 0 & 0 & Q & 26 & 26 & e & s & 0.0157 \\ 
        41 & 581.771556(097) & 581.77210 & ~ & -0.000544 & ~ & 17.4410725 & 0 & 3 & 1 & 0 & 0 & 2 & 0 & 0 & R & 1 & 0 & e & s & 0.0021 \\ 
        41 & 581.806173(105) & 581.80751 & ~ & -0.001337 & ~ & 17.4421103 & 0 & 4 & 4 & 0 & 0 & 3 & 3 & 0 & P & 8 & 9 & f & o & 0.0002 \\ 
        41 & 581.806173(105) & 581.80771 & ~ & -0.001537 & ~ & 17.4421103 & 0 & 4 & 4 & 0 & 0 & 3 & 3 & 0 & P & 8 & 9 & e & o & 0.0002 \\ 
        41 & 581.824473(096) & 581.82494 & ~ & -0.000467 & ~ & 17.4426589 & 0 & 3 & 1 & 0 & 0 & 2 & 0 & 0 & Q & 27 & 27 & e & s & 0.0149 \\ 
        44 & 581.852146(098) & 581.85230 & ~ & -0.000154 & ~ & 17.4434885 & 0 & 1 & 1 & 0 & 0 & 0 & 0 & 0 & P & 2 & 3 & e & s & 0.0003 \\ 
        41 & 581.888125(099) & 581.88838 & 581.888602 & -0.000255 & -0.000477 & 17.4445671 & 0 & 2 & 0 & 0 & 0 & 1 & 1 & 0 & R & 3 & 2 & e & s & 0.0256 \\ 
        41 & 581.891909(099) & 581.89222 & ~ & -0.000311 & ~ & 17.4446806 & 0 & 3 & 1 & 0 & 0 & 2 & 0 & 0 & Q & 28 & 28 & e & s & 0.0146 \\ 
        41 & 581.927927(106) & 581.92818 & ~ & -0.000253 & ~ & 17.4457604 & 0 & 4 & 2 & 0 & 0 & 3 & 1 & 0 & Q & 29 & 29 & f & o & 0.0007 \\ 
        41 & 581.927927(106) & 581.92841 & ~ & -0.000483 & ~ & 17.4457604 & 0 & 4 & 2 & 0 & 0 & 3 & 1 & 0 & Q & 28 & 28 & f & o & 0.0007 \\ 
        41 & 581.929121(111) & 581.92910 & ~ & 0.000021 & ~ & 17.4457962 & 0 & 4 & 2 & 0 & 0 & 3 & 1 & 0 & Q & 30 & 30 & f & o & 0.0010 \\ 
        41 & 581.929121(111) & 581.92969 & ~ & -0.000569 & ~ & 17.4457962 & 0 & 4 & 2 & 0 & 0 & 3 & 1 & 0 & Q & 27 & 27 & f & o & 0.0010 \\ 
        41 & 581.931220(107) & 581.93126 & ~ & -0.000040 & ~ & 17.4458591 & 0 & 4 & 2 & 0 & 0 & 3 & 1 & 0 & Q & 31 & 31 & f & o & 0.0027 \\ 
        41 & 581.931220(107) & 581.93192 & ~ & -0.000700 & ~ & 17.4458591 & 0 & 4 & 2 & 0 & 0 & 3 & 1 & 0 & Q & 26 & 26 & f & o & 0.0027 \\ 
        41 & 581.934474(106) & 581.93477 & ~ & -0.000296 & ~ & 17.4459566 & 0 & 4 & 2 & 0 & 0 & 3 & 1 & 0 & Q & 32 & 32 & f & o & 0.0031 \\ 
        41 & 581.934474(106) & 581.93501 & ~ & -0.000536 & ~ & 17.4459566 & 0 & 4 & 2 & 0 & 0 & 3 & 1 & 0 & Q & 25 & 25 & f & o & 0.0031 \\ 
        41 & 581.938744(109) & 581.93886 & ~ & -0.000116 & ~ & 17.4460847 & 0 & 4 & 2 & 0 & 0 & 3 & 1 & 0 & Q & 24 & 24 & f & o & 0.0013 \\ 
        41 & 581.938744(109) & 581.93972 & ~ & -0.000976 & ~ & 17.4460847 & 0 & 4 & 2 & 0 & 0 & 3 & 1 & 0 & Q & 33 & 33 & f & o & 0.0013 \\ 
        41 & 581.943147(109) & 581.94338 & ~ & -0.000233 & ~ & 17.4462167 & 0 & 4 & 2 & 0 & 0 & 3 & 1 & 0 & Q & 23 & 23 & f & s & 0.0015 \\ 
        41 & 581.945491(109) & 581.94620 & ~ & -0.000709 & ~ & 17.4462869 & 0 & 4 & 2 & 0 & 0 & 3 & 1 & 0 & Q & 34 & 34 & f & o & 0.0009 \\ 
        41 & 581.945491(109) & 581.94731 & ~ & -0.001819 & ~ & 17.4462869 & 0 & 4 & 2 & 0 & 0 & 3 & 3 & 0 & R & 21 & 20 & e & o & 0.0009 \\ 
        41 & 581.948281(109) & 581.94848 & ~ & -0.000199 & ~ & 17.4463705 & 0 & 4 & 2 & 0 & 0 & 3 & 1 & 0 & Q & 22 & 22 & f & s & 0.0005 \\ 
        41 & 581.953805(138) & 581.95406 & ~ & -0.000255 & ~ & 17.4465362 & 0 & 4 & 2 & 0 & 0 & 3 & 1 & 0 & Q & 21 & 21 & f & o & 0.0007 \\ 
        41 & 581.953805(138) & 581.95432 & ~ & -0.000515 & ~ & 17.4465362 & 0 & 4 & 2 & 0 & 0 & 3 & 1 & 0 & Q & 35 & 35 & f & o & 0.0007 \\ 
        41 & 581.959731(139) & 581.96005 & ~ & -0.000319 & ~ & 17.4467138 & 0 & 4 & 2 & 0 & 0 & 3 & 1 & 0 & Q & 20 & 20 & f & s & 0.0002 \\ 
        41 & 581.961634(139) & 581.96205 & ~ & -0.000416 & ~ & 17.4467709 & 0 & 3 & 1 & 0 & 0 & 2 & 0 & 0 & Q & 29 & 29 & e & s & 0.0117 \\ 
        41 & 581.966237(140) & 581.96634 & ~ & -0.000103 & ~ & 17.4469089 & 0 & 4 & 2 & 0 & 0 & 3 & 1 & 0 & Q & 19 & 19 & f & s & 0.0023 \\ 
        41 & 581.972770(130) & 581.97287 & ~ & -0.000100 & ~ & 17.4471047 & 0 & 4 & 2 & 0 & 0 & 3 & 1 & 0 & Q & 18 & 18 & f & s & 0.0004 \\ 
        41 & 581.975167(131) & 581.97582 & ~ & -0.000653 & ~ & 17.4471766 & 0 & 4 & 2 & 0 & 0 & 3 & 1 & 0 & Q & 37 & 37 & f & s & 0.0005 \\ 
        41 & 581.979488(131) & 581.97955 & ~ & -0.000062 & ~ & 17.4473061 & 0 & 4 & 2 & 0 & 0 & 3 & 1 & 0 & Q & 17 & 17 & f & s & 0.0005 \\ 
        41 & 581.986243(144) & 581.98631 & ~ & -0.000067 & ~ & 17.4475086 & 0 & 4 & 2 & 0 & 0 & 3 & 1 & 0 & Q & 16 & 16 & f & s & 0.0005 \\ 
        41 & 581.988735(144) & 581.98938 & ~ & -0.000645 & ~ & 17.4475833 & 0 & 4 & 2 & 0 & 0 & 3 & 1 & 0 & Q & 38 & 38 & f & s & 0.0005 \\ 
        41 & 581.993025(144) & 581.99307 & ~ & -0.000045 & ~ & 17.4477120 & 0 & 4 & 2 & 0 & 0 & 3 & 1 & 0 & Q & 15 & 15 & f & s & 0.0005 \\ 
        41 & 581.999609(123) & 581.99976 & ~ & -0.000151 & ~ & 17.4479093 & 0 & 4 & 2 & 0 & 0 & 3 & 1 & 0 & Q & 14 & 14 & f & s & 0.0006 \\ 
        41 & 582.004116(128) & 582.00494 & ~ & -0.000824 & ~ & 17.4480444 & 0 & 4 & 2 & 0 & 0 & 3 & 1 & 0 & Q & 39 & 39 & f & s & 0.0002 \\ 
        41 & 582.006152(124) & 582.00631 & ~ & -0.000158 & ~ & 17.4481055 & 0 & 4 & 2 & 0 & 0 & 3 & 1 & 0 & Q & 13 & 13 & f & s & 0.0006 \\ 
        43 & 582.009828(114) & 582.01011 & ~ & -0.000282 & ~ & 17.4482157 & 0 & 2 & 0 & 0 & 0 & 1 & 1 & 0 & R & 9 & 8 & e & o & 0.0012 \\ 
        41 & 582.009828(114) & 582.01031 & ~ & -0.000482 & ~ & 17.4482157 & 0 & 4 & 0 & 0 & 0 & 3 & 1 & 0 & R & 10 & 9 & e & o & 0.0012 \\ 
        41 & 582.012527(114) & 582.01267 & ~ & -0.000143 & ~ & 17.4482966 & 0 & 4 & 2 & 0 & 0 & 3 & 1 & 0 & Q & 12 & 12 & f & s & 0.0011 \\ 
        41 & 582.017092(115) & 582.01719 & ~ & -0.000098 & ~ & 17.4484335 & 1 & 2 & 2 & 0 & 1 & 1 & 1 & 0 & P & 14 & 15 & f & s & 0.0015 \\ 
        41 & 582.018532(114) & 582.01877 & ~ & -0.000238 & ~ & 17.4484766 & 0 & 4 & 2 & 0 & 0 & 3 & 1 & 0 & Q & 11 & 11 & f & s & 0.0008 \\ 
        41 & 582.022284(123) & 582.02257 & ~ & -0.000286 & ~ & 17.4485891 & 0 & 4 & 2 & 0 & 0 & 3 & 1 & 0 & Q & 40 & 40 & f & s & 0.0017 \\ 
        41 & 582.024284(118) & 582.02456 & ~ & -0.000276 & ~ & 17.4486491 & 0 & 4 & 2 & 0 & 0 & 3 & 1 & 0 & Q & 10 & 10 & f & s & 0.0008 \\ 
        41 & 582.029921(118) & 582.02998 & ~ & -0.000059 & ~ & 17.4488181 & 0 & 4 & 2 & 0 & 0 & 3 & 1 & 0 & Q & 9 & 9 & f & s & 0.0005 \\ 
        41 & 582.034291(117) & 582.03446 & ~ & -0.000169 & ~ & 17.4489491 & 0 & 3 & 1 & 0 & 0 & 2 & 0 & 0 & Q & 30 & 30 & e & s & 0.0122 \\ 
        41 & 582.039516(105) & 582.03957 & ~ & -0.000054 & ~ & 17.4491057 & 0 & 4 & 2 & 0 & 0 & 3 & 1 & 0 & Q & 7 & 7 & f & s & 0.0004 \\ 
        41 & 582.042234(107) & 582.04236 & ~ & -0.000126 & ~ & 17.4491872 & 0 & 4 & 2 & 0 & 0 & 3 & 1 & 0 & Q & 41 & 41 & f & s & 0.0008 \\ 
        41 & 582.043426(107) & 582.04365 & ~ & -0.000224 & ~ & 17.4492229 & 0 & 4 & 2 & 0 & 0 & 3 & 1 & 0 & Q & 6 & 6 & f & s & 0.0014 \\ 
        41 & 582.046956(107) & 582.04721 & ~ & -0.000254 & ~ & 17.4493287 & 0 & 4 & 2 & 0 & 0 & 3 & 1 & 0 & Q & 5 & 5 & f & s & 0.0018 \\ 
        41 & 582.050008(111) & 582.05022 & ~ & -0.000212 & ~ & 17.4494203 & 0 & 4 & 2 & 0 & 0 & 3 & 1 & 0 & Q & 4 & 4 & f & s & 0.0022 \\ 
        41 & 582.052152(113) & 582.05266 & ~ & -0.000508 & ~ & 17.4494845 & 0 & 4 & 2 & 0 & 0 & 3 & 1 & 0 & Q & 3 & 3 & f & s & 0.0014 \\ 
        41 & 582.054220(113) & 582.05450 & ~ & -0.000280 & ~ & 17.4495465 & 0 & 4 & 2 & 0 & 0 & 3 & 1 & 0 & Q & 2 & 2 & f & s & 0.0002 \\ 
        41 & 582.063118(105) & 582.06321 & 582.063199 & -0.000092 & -0.000081 & 17.4498133 & 1 & 1 & 1 & 0 & 1 & 0 & 0 & 0 & P & 15 & 16 & e & s & 0.0025 \\ 
        41 & 582.073253(104) & 582.07327 & 582.073253 & -0.000014 & 0.000000 & 17.4501171 & 0 & 1 & 1 & 0 & 0 & 0 & 0 & 0 & P & 7 & 8 & e & s & 1.3931 \\ 
        41 & 582.079876(155) & 582.07998 & ~ & -0.000104 & ~ & 17.4503157 & 0 & 4 & 2 & 0 & 0 & 3 & 1 & 0 & Q & 4 & 4 & e & s & 0.0011 \\ 
        41 & 582.091695(137) & 582.09186 & ~ & -0.000165 & ~ & 17.4506700 & 0 & 4 & 2 & 0 & 0 & 3 & 1 & 0 & Q & 5 & 5 & e & s & 0.0006 \\ 
        41 & 582.105832(102) & 582.10612 & ~ & -0.000288 & ~ & 17.4510938 & 0 & 4 & 2 & 0 & 0 & 3 & 1 & 0 & Q & 6 & 6 & e & s & 0.0014 \\ 
        41 & 582.109215(101) & 582.10945 & ~ & -0.000235 & ~ & 17.4511952 & 0 & 3 & 1 & 0 & 0 & 2 & 0 & 0 & Q & 31 & 31 & e & s & 0.0128 \\ 
        41 & 582.122526(103) & 582.12275 & ~ & -0.000224 & ~ & 17.4515943 & 0 & 4 & 2 & 0 & 0 & 3 & 1 & 0 & Q & 7 & 7 & e & s & 0.0003 \\ 
        41 & 582.141047(105) & 582.14124 & ~ & -0.000193 & ~ & 17.4521495 & 0 & 3 & 1 & 0 & 0 & 2 & 2 & 0 & R & 13 & 12 & e & s & 0.0016 \\ 
        42 & 582.145835(105) & 582.14614 & ~ & -0.000305 & ~ & 17.4522931 & 0 & 1 & 1 & 0 & 0 & 0 & 0 & 0 & R & 8 & 7 & e & s & 0.0052 \\ 
        41 & 582.162889(108) & 582.16313 & ~ & -0.000241 & ~ & 17.4528043 & 0 & 4 & 2 & 0 & 0 & 3 & 1 & 0 & Q & 9 & 9 & e & s & 0.0004 \\ 
        41 & 582.186785(112) & 582.18703 & ~ & -0.000245 & ~ & 17.4535207 & 0 & 3 & 1 & 0 & 0 & 2 & 0 & 0 & Q & 32 & 32 & e & s & 0.0123 \\ 
        41 & 582.212747(114) & 582.21299 & ~ & -0.000243 & ~ & 17.4542990 & 0 & 4 & 2 & 0 & 0 & 3 & 1 & 0 & Q & 11 & 11 & e & s & 0.0007 \\ 
        41 & 582.241277(116) & 582.24147 & ~ & -0.000193 & ~ & 17.4551544 & 0 & 4 & 2 & 0 & 0 & 3 & 1 & 0 & Q & 12 & 12 & e & s & 0.0011 \\ 
        41 & 582.265908(108) & 582.26640 & ~ & -0.000492 & ~ & 17.4558928 & 0 & 2 & 2 & 0 & 0 & 1 & 1 & 0 & P & 7 & 8 & f & s & 0.0603 \\ 
        41 & 582.266871(109) & 582.26721 & ~ & -0.000339 & ~ & 17.4559216 & 0 & 3 & 1 & 0 & 0 & 2 & 0 & 0 & Q & 33 & 33 & e & s & 0.0066 \\ 
        41 & 582.271872(110) & 582.27232 & ~ & -0.000448 & ~ & 17.4560716 & 0 & 4 & 2 & 0 & 0 & 3 & 1 & 0 & Q & 13 & 13 & e & s & 0.0040 \\ 
        41 & 582.305078(114) & 582.30553 & ~ & -0.000452 & ~ & 17.4570671 & 0 & 4 & 2 & 0 & 0 & 3 & 1 & 0 & Q & 14 & 14 & e & s & 0.0015 \\ 
        41 & 582.323201(132) & 582.32355 & ~ & -0.000349 & ~ & 17.4576104 & 0 & 2 & 2 & 0 & 0 & 1 & 1 & 0 & P & 7 & 8 & e & s & 0.0553 \\ 
        41 & 582.340790(138) & 582.34111 & ~ & -0.000320 & ~ & 17.4581377 & 0 & 4 & 2 & 0 & 0 & 3 & 1 & 0 & Q & 15 & 15 & e & s & 0.0004 \\ 
        41 & 582.349705(134) & 582.35002 & ~ & -0.000315 & ~ & 17.4584049 & 0 & 3 & 1 & 0 & 0 & 2 & 0 & 0 & Q & 34 & 34 & e & s & 0.0037 \\ 
        42 & 582.369249(180) & 582.36954 & ~ & -0.000291 & ~ & 17.4589909 & 0 & 2 & 2 & 0 & 0 & 1 & 1 & 0 & R & 8 & 7 & e & s & 0.0002 \\ 
        41 & 582.378885(177) & 582.37905 & ~ & -0.000165 & ~ & 17.4592797 & 0 & 4 & 2 & 0 & 0 & 3 & 1 & 0 & Q & 16 & 16 & e & s & 0.0004 \\ 
        41 & 582.414368(150) & 582.41438 & ~ & -0.000012 & ~ & 17.4603435 & 0 & 3 & 1 & 0 & 0 & 2 & 2 & 0 & R & 13 & 12 & f & s & 0.0011 \\ 
        41 & 582.419100(152) & 582.41935 & ~ & -0.000250 & ~ & 17.4604854 & 0 & 4 & 2 & 0 & 0 & 3 & 1 & 0 & Q & 17 & 17 & e & s & 0.0011 \\ 
        41 & 582.435321(153) & 582.43544 & ~ & -0.000119 & ~ & 17.4609716 & 0 & 3 & 1 & 0 & 0 & 2 & 0 & 0 & Q & 35 & 35 & e & s & 0.0025 \\ 
        41 & 582.461529(138) & 582.46200 & ~ & -0.000471 & ~ & 17.4617573 & 0 & 4 & 2 & 0 & 0 & 3 & 1 & 0 & Q & 18 & 18 & e & s & 0.0002 \\ 
        41 & 582.475740(134) & 582.47616 & ~ & -0.000420 & ~ & 17.4621834 & 0 & 3 & 3 & 0 & 0 & 2 & 2 & 0 & P & 7 & 8 & e & o & 0.0027 \\ 
        41 & 582.475740(134) & 582.47649 & ~ & -0.000750 & ~ & 17.4621834 & 0 & 3 & 3 & 0 & 0 & 2 & 2 & 0 & P & 7 & 8 & f & o & 0.0027 \\ \hline
\caption{A complete table of measured N$_2$O transitions. The columns are as follows. iso: a code used in HITRAN~\citep{Gordon2022} to identify the isotope of N$_2$O; 41 means $^{14}$N$_2^{16}$O, 42 means $^{14}$N$^{15}$N$^{16}$O, 43 means $^{15}$N$^{14}$N$^{16}$O, 44 means $^{14}$N$_2^{18}$O and 45 means $^{14}$N$_2^{17}$O. wn (observed): our measured wavenumbers (in cm$^{-1}$) with 2$\sigma$ uncertainties given in parentheses. wn (hi): transition wavenumbers listed in HITRAN (in cm$^{-1}$). wn (nist): transition wavenumbers listed as calibration lines by NIST (in cm$^{-1}$). `o-ht': difference between our measured wavenumbers and those in HITRAN (in cm$^{-1}$). o-nist: difference between our measured wavenumbers and those in NIST database (in cm$^{-1}$). $f$ (observed): our measured frequency (in THz). $v_1',v_2',l',v_3'$: vibrational quantum numbers of upper state. $v_1,v_2,l,v_3$: vibrational quantum numbers of lower state. PQR: type of rotational transition. $J'$: rotational quantum number of upper state. $J$: rotational quantum number of lower state. p: e/f symmetry of the lower level. ov: whether the line is overlapped with other lines; s means no overlap and o means overlap. amp: fitted amplitude value using Voigt line profile.}
    \label{tab:n2odata}
    \end{longtable}

\end{landscape}\restoregeometry
\clearpage

\end{document}